\documentclass[%
reprint,
superscriptaddress,
prb,
floatfix,
]{revtex4-2}

\usepackage{graphicx}
\usepackage{dcolumn}
\usepackage{bm}
\usepackage{hyperref}
\hypersetup{
    colorlinks=true,
    linkcolor=red, citecolor=blue,
    filecolor=blue,      
    urlcolor=blue,
    pdftitle={KMgX},
    }
\usepackage{comment}

\usepackage{soul}
\usepackage{multirow}
\usepackage{hhline}

\begin{document}
\title{Effect of hydrostatic pressure and alloying on thermoelectric properties of van der Waals solid KMgSb: An \textit{ab-initio} study}

\author{Vikrant~Chaudhary}
\affiliation{Department of Chemistry, Indian Institute of Technology Roorkee, Roorkee, Uttarakhand, India-247667}
\author{Tulika~Maitra}
\affiliation{Department of Physics, Indian Institute of Technology Roorkee, Roorkee, Uttarakhand, India-247667}
\author{Tashi~Nautiyal}
\affiliation{Department of Physics, Indian Institute of Technology Roorkee, Roorkee, Uttarakhand, India-247667}
\author{Jeroen~van~den~Brink}
\affiliation{Institute for Theoretical Solid State Physics, IFW Dresden, Helmholtzstrasse 20, 01069 Dresden, Germany}
\altaffiliation[Also at ]{Institute for Theoretical Physics and WÃ¼rzburg-Dresden Cluster of Excellence ct.qmat, Technische UniversitÃ¤t Dresden, 01069 Dresden, Germany}
\author{Hem~C.~Kandpal}\email{Corresponding author: hem.kandpal[at]cy.iitr.ac.in}
\affiliation{Department of Chemistry, Indian Institute of Technology Roorkee, Roorkee, Uttarakhand, India-247667}

\date{\today}
	
\begin{abstract}
Through a combined first-principles and Boltzmann transport theory, we systematically investigate the thermal and electrical transport properties of the unexplored ternary quasi two-dimensional KMgSb system of KMgX (X = P, As, Sb, and Bi) family. Herein, the transport properties of KMgSb under the application of hydrostatic pressure and alloy engineering are reported.  At a carrier concentration of $\sim8\times10^{19}~\mathrm{cm^{-3}}$, the figure of merit zT ($\sim0.75$) for both the $n$-type and $p$-type of KMgSb closely matched, making it an attractive option for engineering both legs of a thermoelectric device using the same material. This is particularly desirable for high-performance thermoelectric applications. Furthermore, the zT value increases as pressure decreases, further enhancing its potential for use in thermoelectric devices. In the case of substitutional doping (replacing 50 \% Sb by Bi atom), we observed $\sim49~\%$ (in-plane) increase in the peak thermoelectric figure of merit (zT). The maximum zT value obtained after alloy engineering is $\sim1.45$ at 900~K temperature. Hydrostatic pressure is observed to be a great tool to tune the lattice thermal conductivity ($\kappa_L$). We observed that the negative pressure-like effects could be achieved by chemically doping bigger-size atoms, especially when $\kappa_L$ is a property under investigation. Through our computational investigation, we explain that hydrostatic pressure and alloy engineering may improve thermoelectric performance dramatically.
\end{abstract}

\maketitle
\section{\label{sec:Intro}Introduction}
Thermometric materials have been around for many decades now. Many materials have been developed and identified for thermoelectric devices-based applications \cite{Rev1, Rev2,ref1, ref2, ref3,ref4, 1PbTe1, ref6, ref7}. The performance and efficiency of thermoelectric materials are judged using a dimensionless figure of merit $zT = S^{2}\sigma T/(\kappa_{e}+\kappa_{L})$, where $S$, $\sigma$, $\kappa_{e}$, and $\kappa_{L}$ are Seebeck coefficient, electrical conductivity, electronic thermal conductivity, and lattice thermal conductivity, respectively. Among the many promising materials \cite{1PbTe1,2SnSe1,3SnSe2,4SnTe,5GeTe,6SiGe,7HH1,8HH2,9HH3,10HH4}, the materials that are layered in nature, like SnSe, show high zT value due to their low thermal conductivity \cite{Rev2}. Thus, the quasi-2D nature of layered materials leads to improved thermoelectric performance, and such materials are widely used in thermoelectric devices. SnSe is an example of a layered material where high zT values have been reported in $n$-type as well as $p$-type doping regions. The layered SnSe has shown exemplary thermoelectric performance in both, single crystals and polycrystalline phases \cite{2022_SnSe}. With the same motive, we have studied the less explored KMgSb from the KMgX (X =  P, As, Sb, and Bi) family, where the compounds have a quasi-2D type of crystal structure. As per our knowledge, all the members of the KMgX family, except KMgN, have been experimentally realized \cite{Sc1979}. 
	
The topological \cite{KMgBi3} and thermoelectric \cite{KMgBi2} properties of KMgBi are well explored among all members in the family. KMgBi has been reported to host type-I Dirac points \cite{KMgBi3} and in another study was reported as a narrow band gap ($\sim$11.2~meV) semiconductor \cite{KMgBi}. Contrary to this, a density functional theory (DFT) based study has reported that KMgBi has a larger band gap ($\sim$ 280~meV) and shows a high zT value ($\sim 2.21$) with $p$-type doping, which is expected to improve further with alloy engineering \cite{KMgBi2}.
	
Besides KMgBi, a few studies have been reported related to other members of the KMgX (X =  P, As, Sb, and Bi) family in hexagonal, orthorhombic, and half-Heusler phases. The thermoelectric properties of KMgP have been investigated in the half-Heusler thin film \cite{KMgP_HH} and MoS$_{2}$ type 2D structure \cite{KMgP_MoS2_2D}. The ferroelectric and antiferroelectric properties of KMgSb and KMgBi have been reported in the hexagonal \cite{KMgSb_Hexa} and orthorhombic \cite{KMgSb_Otho} phase, respectively. In another study, the optical and elastic properties of KMgX (X =  P, As, Sb, and Bi) have been explored in a hypothetical half-Heusler structure \cite{KMgX_HH}. Apart from these, no thermal and electronic transport studies have been reported, as per our knowledge, in the above-mentioned or experimentally realized tetragonal ($P4/nmm$) structures. Thus, we have investigated the thermoelectric transport of KMgSb from the KMgX family in the less studied and experimentally realized quasi-2D tetragonal structure\cite{Sc1979}. We find that as we go down in the KMgX family, the band gap lowers, and this tunability can play important role in getting a good thermoelectric performance. The known studies on KMgBi \cite{KMgBi2,KMgBi3} further motivated us to investigate KMgSb in quasi-2D structure.
	
Over the time, many techniques have been used to enhance the thermoelectric performance of materials. For example, PbTe and PbTe-based materials have been developed as high-performance thermoelectric materials by optimizing phonon and charge transport. These techniques include tuning the Fermi level by changing carrier density, increasing effective mass by band structure engineering, decreasing lattice thermal conductivity by nanostructuring, etc. \cite{PbTe}. Another promising way of improving thermoelectric performance is alloy engineering \cite{2022_Mg3Bi2}. 
	
We picked KMgSb as a candidate material for this study. In this work, we have investigated the effect of hydrostatic pressure and isoelectronic doping (at the Sb sites) on the thermoelectric performance of KMgSb. The hydrostatic pressure results in varying interlayer spacing in the quasi-2D KMgSb, which affects the zT value significantly. As, Bi, and Sb are iso-electronic elements and doping Sb and Bi help only in tuning the volume and in generating creating substitutional defects. Moreover, increased volume by chemical doping can be thought of as a condition created by negative pressure, which is also an important parameter and helps in the enhancement of thermoelectric performance. Further details are given in the result section of this paper. 
	%
	%
\section{\label{sec:Comp}COMPUTATIONAL DETAILS}
The computations performed in this work have three major components, namely (a) DFT for electronic band structure and density functional perturbation theory (DFPT) for mechanical and dielectric properties, (b) phonon or lattice dynamics, and (c) the scattering rates and transport properties. 
	
The DFT and DFPT calculations were performed using the Vienna \textit{Ab-initio} Simulation Package (VASP) \cite{1VASP,2VASP,3VASP,4VASP}. The all-electron projector augmented wave method \cite{5VASP,paw1} was adopted with a plane-wave cutoff energy of 550~eV and a breaking condition of 10$^{-8}$~eV on the self-consistency field (SCF) cycles were used. We have used the generalized gradient approximation (GGA) of J. P. Perdew, K. Burke, and M. Ernzerhof (PBE) \cite{GGA}, revised PBE for solids (PBEsol) \cite{PBEsol}, and HSE06 \cite{HSE06} hybrid functional for the exchange-correlation part. Since KMgX family compounds are layered in nature, the van der Waals density functional (optB86b-vdW) \cite{OptB86b-vdW} is used on top of the PBE and HSE06 functionals. All crystal structures are optimized on a $21\times21\times15$ k-mesh. During the ionic relaxation, the forces were relaxed with a breaking condition of 0.01~eV/\AA\/. Furthermore, the mechanical and dielectric properties were calculated using the DFPT module of VASP, and the results were used for scattering rate evaluation.
	
The phonon band structures and Gr\"uneisen parameters were obtained using the harmonic approximation in the Phonopy package \cite{phonopy} and DFPT module of VASP. For Gr\"uneisen parameter calculations, the volume of the unit cell was changed by $\sim\pm1.0~\%$. We obtained the lattice thermal conductivity ($\kappa_{L}$) using Phono3py \cite{phono3py}, and ShengBTE \cite{ShengBTE_2014,thirdorder} packages on top of the DFT results from VASP. A $2\times2\times2$ supercell was used to calculate the third-order \cite{thirdorder} and fourth-order \cite{2016_Four_Phonon,2017_Four_Phonon,2022_Four_Phonon} anharmonic inter-atomic force constants (IFCs). The second-order IFCs were obtained using Phonopy by generating large supercell ($3\times3\times2$) structures. Phono3py gives results within the relaxation time approximation (RTA), whereas ShengBTE allows us to calculate $\kappa_L$ using RTA as well as the iterative method. We observed a close agreement between the $\kappa_L$ obtained from both methods. The compounds studied in this work are ionic in nature. Thus, we included non-analytical corrections in our calculations.
	
The last component in computations is related to the scattering rates and transport properties, which are calculated using AMSET code \cite{AMSET}. AMSET takes mechanical, piezoelectric, and dielectric properties, wave functions, band structure, and deformation potential in input, which we obtained from the VASP calculations. The electronic and thermal transport properties are obtained by numerically solving the linearized Boltzmann transport equations (BTEs). For a very long time, these BTEs were solved within the constant relaxation time approach (CRTA) due to the complex nature of the scattering processes. In AMSET, the scattering rates are obtained using Fermi's golden rule,
	\begin{equation}
		\tau_{i\rightarrow f}^{-1}=\frac{2\pi}{\hbar} |g_{fi}(\textbf{k},\textbf{q})|^{2}\delta(\epsilon_{i}-\epsilon_{f}) \label{Fermi}
	\end{equation}
where $i$ is the initial and $f$ is the final state, $\tau$ is the relaxation time, $g$ is the coupling matrix, and $\epsilon_i$ ($\epsilon_f$) is the initial (final) energy of the electron. The $g$ matrices are the scattering rate calculations' most significant and complex part. These matrices were obtained using acoustic deformation potential (ADP), ionized impurity (IMP), and polar optical phonon (POP) scattering mechanisms in this work. The ADP and IMP are elastic processes, while POP is inelastic \cite{AMSET}. The relaxation time obtained is automatically used in various transport properties, which are obtained by solving the Boltzmann transport equations module of the AMSET. Various transport coefficients are obtained from the generalized transport equation
	\begin{equation}
		\mathcal{L}^{\alpha}(\mu, T)=q^{2}\int \Sigma(\epsilon)~(\epsilon-\mu)^{\alpha}\left(-\frac{\partial f^{0}(\epsilon, T)}{\partial \epsilon}\right) d\epsilon \label{Onsager}
	\end{equation}
	where, $q$, $\Sigma(\epsilon)$, $\mu$ and $f^{0}(\epsilon, T)$ are electronic charge, spectral conductivity, chemical potential, and Fermi-Dirac distribution function, respectively \cite{btp2,onsager,AMSET}. The Seebeck coefficient (S), electronic conductivity ($\sigma$), and electronic component of thermal conductivity ($\kappa_e$), are given by $\frac{1}{qT} \frac{\mathcal{L}^{1}}{\mathcal{L}^{0}}$, $\mathcal{L}^{0}$, and $\frac{1}{q^2T} \left[\frac{(\mathcal{L}^{1})^2}{\mathcal{L}^{0}} -\mathcal{L}^2\right]$, respectively. 
	%
\begin{table*}[t]
	\caption{\label{tab:T1} The lattice parameters of the KMgX (X = P, As, Sb, and Bi)\textsuperscript{\emph{a}} family members and related alloys. The values in parentheses are relative percentage errors with respect to the experimental lattice constants. \textsuperscript{\emph{a}}KMgN is not experimentally known, hence not included in the study. \textsuperscript{\emph{b}}With optB86b-vdW corrections. NA stands for Not Available.}
 \begin{ruledtabular}
      \begin{tabular}{@{}lcllllll@{}}
			\multirow{2}{*}{Composition} & \multirow{2}{*}{} & \multicolumn{6}{c}{Lattice parameter (\AA)} \\
			& & PBE & PBEsol & PBE\textsuperscript{\emph{b}} & HSE06 & HSE06\textsuperscript{\emph{b}} & Expt. \cite{Sc1979} \\
			\hline
			\multirow{2}{*}{KMgP} & a & 4.462 (0.36) &4.426 (-0.45) & 4.426 (-0.45) & 4.436 (-0.22) & 4.428 (-0.40) & 4.446 \\
			& c & 7.599 (0.73) &7.526 (-0.24) & 7.532 (-0.16) & 7.605 (0.81) & 7.536 (-0.11) & 7.544 \\
			\multirow{2}{*}{KMgAs} & a & 4.577 (0.68) & 4.526 (-0.44) & 4.530 (-0.35) & 4.546 (0.00) & 4.536 (-0.22) & 4.546\\
			& c & 7.872 (2.02) & 7.736 (0.26) & 7.742 (0.34) & 7.796 (1.04) & 7.738 (0.28) & 7.716 \\
			\multirow{2}{*}{KMgSb} & a & 4.840 (0.58) & 4.781 (-0.64) & 4.794 (-0.37) & 4.803 (-0.19) & 4.794 (-0.37) & 4.812 \\
			& c & 8.341 (1.69)& 8.223 (0.26)& 8.221 (0.23)& 8.258 (0.68)& 8.204 (0.02) & 8.202 \\
			\multirow{2}{*}{KMgSb$_{0.5}$Bi$_{0.5}$} & a & 4.888 & 4.820 & 4.834 & 4.848 &4.839 & \multirow{2}{*}{NA}  \\
			& c & 8.446 & 8.311 & 8.306 & 8.355 & 8.288 \\
			\multirow{2}{*}{KMgBi} & a & 4.933 (1.06) & 4.870 (-0.22) & 4.882 (0.02)& 4.896 (0.31)& 4.885 (0.08) & 4.881 \\
			& c & 8.545 (1.94)& 8.397 (0.18)& 8.382 (0.00)& 8.452 (0.84)& 8.365 (-0.20)& 8.382 \\

		\end{tabular}\\	
  \end{ruledtabular}
\end{table*}
\section{\label{sec:Res}Results and Discussion}
\subsection{\label{subsec:Struct}Structural analysis}
\begin{figure}[b]
		\centering\includegraphics[scale=0.33]{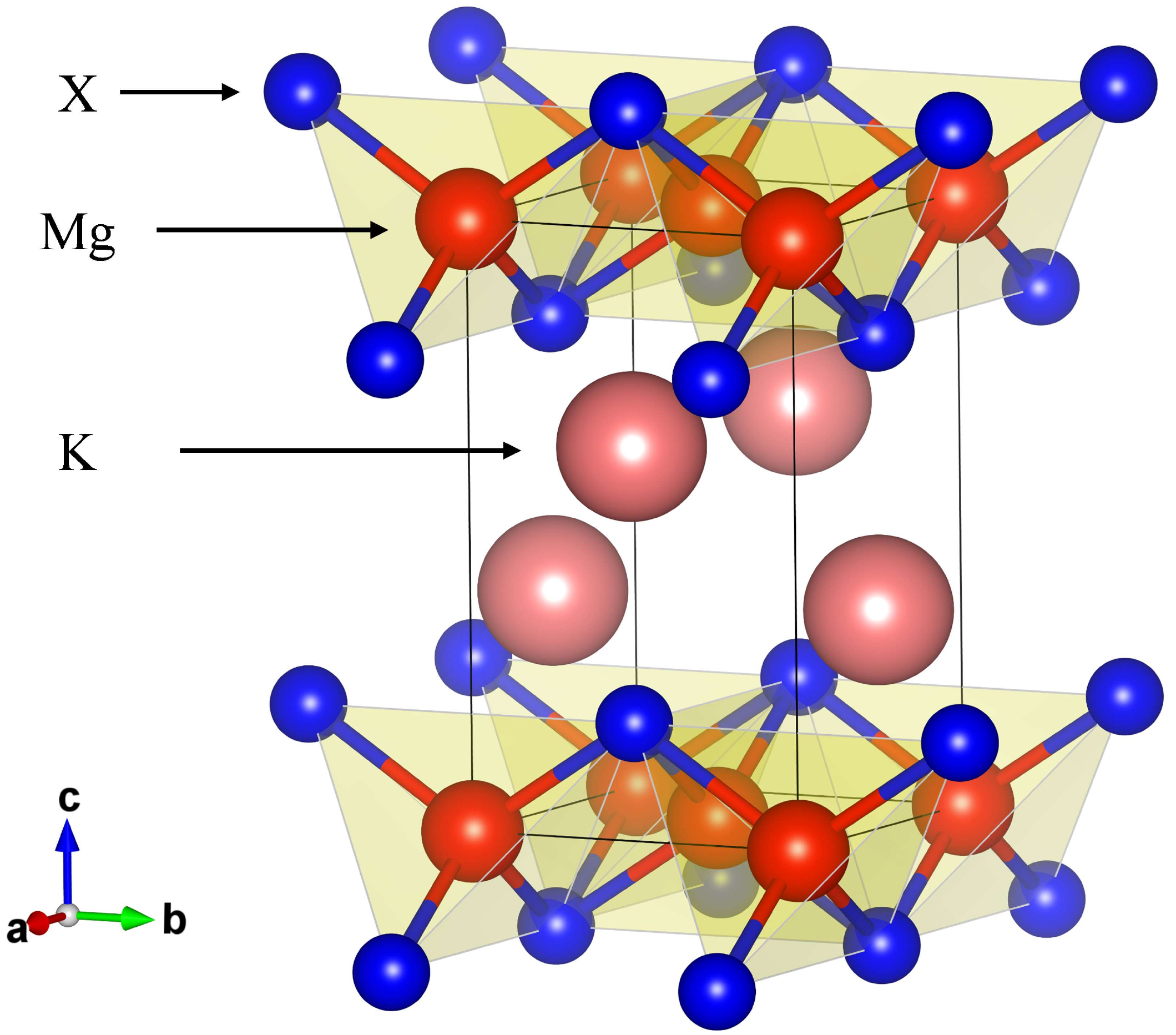}
		\caption{\label{fig:structure}The crystal structure of KMgX (X = P, As, Sb, and Bi) in $P4/nmm$ space group.}
	\end{figure}

The compounds of KMgX family (X = P, As, Sb, and Bi) crystallize in $P4/nmm$ space group (No. 129) \cite{Sc1979}. The K, Mg, and X occupy $2c$ (0, 0.5, $\sim$ 0.65), $2a$ (0, 0, 0), and $2c$ (0, 0.5, $\sim$ 0.20) sites, respectively. The $z$ component of the sites occupied by K and X are free positions, which we have relaxed during the structural optimization. The structures of KMgX are layered in the $c$ direction, consisting of alternating K and Mg-X layers. In the Mg-X layer, Mg and X atoms form edge-sharing tetrahedra, as shown in Fig.~\ref{fig:structure}. Since K is an alkali metal, it tends to give an electron to the MgX tetrahedron, resulting in an electrostatic interaction between K$^{+}$ and [MgX]$^{-}$. This electrostatic interaction is weaker than the covalent bond between Mg and X atoms due to the large separation between K and Mg-X layers.  
	%
	%
	\begin{table}[b]
		\centering\caption{\label{tab:T2} The band gap of the KMgX (X = P, As, Sb, and Bi)\textsuperscript{\emph{a}} family compounds and related alloys. The band gap in parentheses is obtained by including spin-orbit coupling (SOC) in calculations. The experimental band gap of KMgBi is $\sim0.011~$eV \cite{KMgBi}. \textsuperscript{\emph{b}}With optB86b-vdW corrections.}
  \begin{ruledtabular}
		\begin{tabular}{@{}lcccc@{}}
	
			\multirow{2}{*}{Composition} & \multicolumn{4}{c}{Band Gap E$_g$ (eV)} \\
			& PBE & PBE\textsuperscript{\emph{b}} & HSE06 & HSE06\textsuperscript{\emph{b}}\\
			\hline
			KMgP & 1.70 (1.63) &  1.80 (2.02) & 2.49 (2.47) &2.55 (2.78)\\
			KMgAs &  1.12 (1.04) & 1.25 (1.39) & 1.92 (1.83) & 1.96 (2.11)\\
			KMgSb & 1.26 (1.07) & 1.34 (1.25) & 1.93 (1.74) & 1.96 (1.87) \\
			KMgSb$_{0.5}$Bi$_{0.5}$  & 0.76 (0.38) &  0.89 (0.69)  & 1.43 (1.00) & 1.50 (1.27)\\
			KMgBi & 0.31 (0.00) & 0.49 (0.16) & 0.93 (0.35) & 1.03 (0.64)\\

		\end{tabular}\\      
\end{ruledtabular}

\end{table}

 \begin{figure*}[t]
		\centering\includegraphics[scale=0.58]{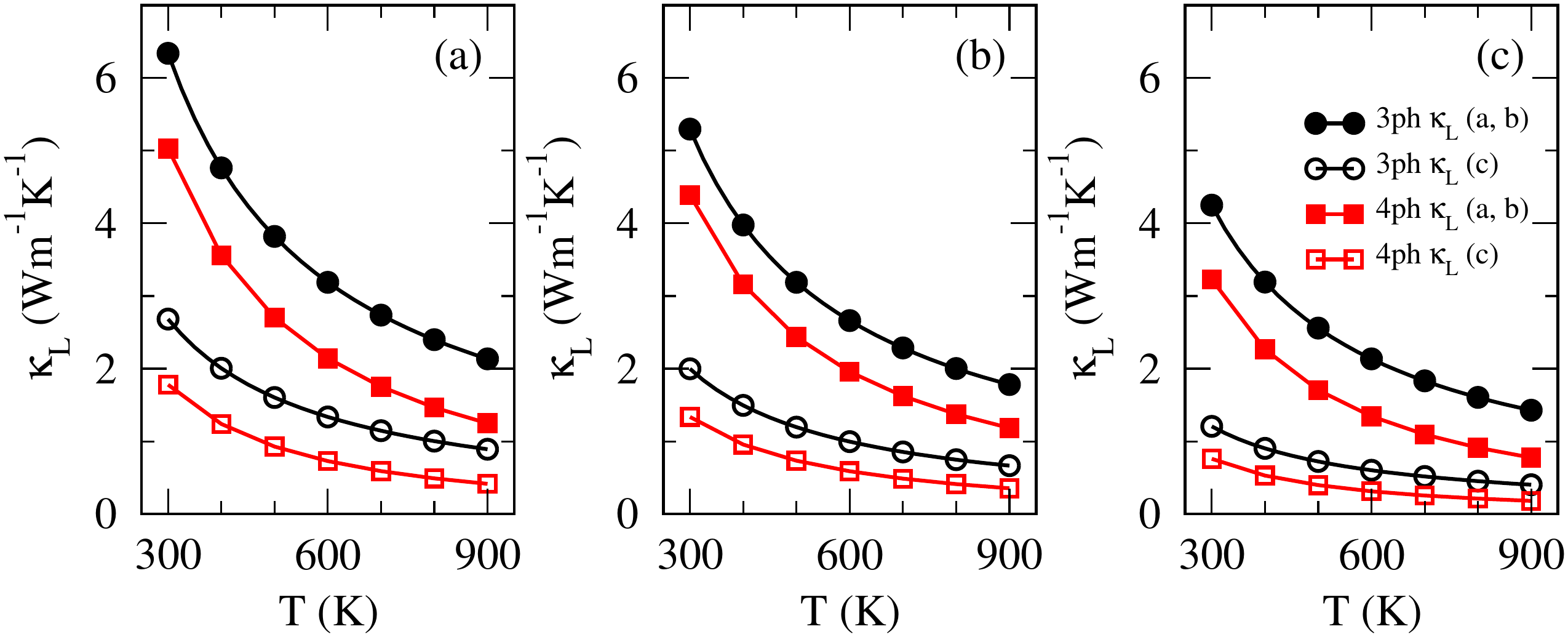}
		\caption{\label{fig:ltc}The lattice thermal conductivity $\kappa_L$ of KMgSb at (a) 1.0~GPa, (b) ambient, and (c) -1.0~GPa pressure. The black curves show the $\kappa_L$ calculated using $3^{rd}$ order IFCs and red curves represents values obtained using $4^{th}$ and $3^{rd}$ order IFCs.}
\end{figure*}

\begin{figure*}[t]
		\centering\includegraphics[scale=0.57]{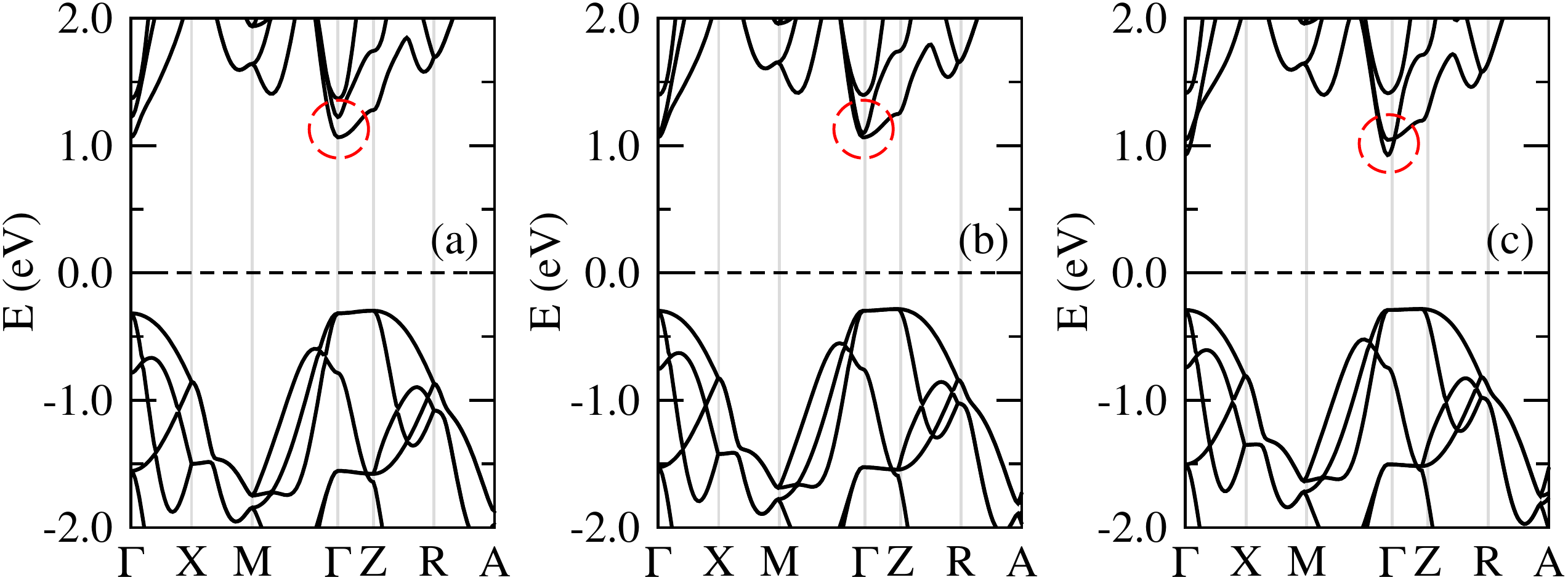}
		\caption{\label{fig:band}The electronic band structures of KMgSb with van der Waals corrections and at (a) 1.0~GPa, (b) ambient, and (c) -1.0~GPa pressure.}
\end{figure*}  
The interaction between the layers of KMgX is van der Waals in nature. Therefore, it becomes inevitable to exclude van der Waals corrections in DFT calculations. The unit cell parameters were fully relaxed using PBE, PBEsol, PBE with optB86b-vdW, HSE06, and HSE06 with optB86b-vdW exchange-correlation functionals, as given in Table \ref{tab:T1}. The lattice parameters obtained with the optB86b-vdW correction and using the HSE06 and PBE functionals are in excellent agreement with the experimental values. As we go down in the table, PBE with optB8b-vdW gives a more accurate value, and the most accurate results are obtained for KMgBi. The PBE (with optB86b-vdW) calculated $a$ lattice constant of KMgBi differs only by $0.001$ \AA~ from the experimental value and $c$ matches exactly up to $3$ decimal points as given in Table \ref{tab:T1}. Without the van der Waals corrections, the $c$ lattice parameter differs significantly from the experimental value. 

The band gap is another parameter that we have tested with different exchange-correlation functionals. We only have the experimental band gap ($\sim$11.2~meV) for KMgBi, obtained from resistivity data measured between 40~K and 100~K \cite{KMgBi}. The band gap value obtained from PBE with optB86b-vdW seems to be the best prediction for KMgBi, considering many approximations that DFT is based on. The band gap values of various compositions are given in Table \ref{tab:T2}. 

We thoroughly scanned the KMgX family compounds and found that KMgSb is a good candidate for pressure study and alloy engineering. KMgBi is reported to show good thermoelectric performance \cite{KMgBi2}. However, the results published on KMgBi are based on PBE optimized crystal structure, which lacks the inclusion of van der Waal's correction mandatory for this family of compounds. The PBE optimized lattice constants of KMgBi shows large deviation from experimental values, as shown in table \ref{tab:T1}. Also, the transport properties of KMgBi reported in earlier work has been obtained by scaling the PBE bandgap to the HSE06+GW+SOC bandgap, keeping the band curvature fixed. This approach should not reflect the effect of HSE06 and GW in the final transport results. Also, the bandgap of KMgBi obtained using HSE06+GW+SOC is overestimated ($\sim$280~meV)~\cite{KMgBi2}. The bandgap of KMgBi obtained in our work with GGA+optB86b-vdW+SOC is quantitatively the best theoretically predicted value. Also, the predicted lattice constants with PBE+optB86b-vdW are the most accurate (closest to experimental) values. Our approach is to work with the functionals that gives most accurate lattice constants and bandgap values. In light of available studies, we have picked unexplored KMgSb from the family and carried out a detailed investigation for understanding the thermoelectric transport in the KMgX family. In this paper, we selected PBE with optB86b-vdW corrections for all the calculations, including phonon and charge transport evaluation. By alloy engineering (Bi doping in KMgSb in this work), a composition with promising thermoelectric transport properties is reported. The pressure study and alloy engineering (Bi doping in KMgSb) are discussed in the following sections of this paper.

\subsection{\label{subsec:Pressure}KMgSb under hydrostatic pressure}
The KMgSb has a low bulk modulus ($\sim$25~GPa), as we found in our calculations. The low bulk modulus indicates that KMgSb is a low-strength material and large hydrostatic pressure will possibly change the crystal structure of the compound. Herein, we are not interested in a pressure-driven phase transition. Therefore, we have limited our study to -1.0~GPa and 1.0~GPa pressure. To understand the effect of the pressure on the unit cell, the change in lattice parameters of KMgSb and KMgBi with pressure is given in the supplementary information (SI).

The phonon band structure and mode Gr\"uneisen parameter ($\gamma$) are the key tools to understand the various aspects of lattice dynamics, which we calculated using the phonopy package \cite{phonopy}. The mode Gr\"uneisen parameter is a direct measure of the degree of anharmonicity in phonon modes. A larger value of $\gamma$ leads to a low $\kappa_L$, as per the inverse square relation ($\kappa_L \propto 1/\gamma^{2}$) \cite{slack}. The maximum value of $\gamma$ ($\sim$1.75 at ambient pressure) in the acoustic region (below $\sim$2.4~THz) rises to $\sim$2.25 at -1.0 GPa (see SI). We are focusing on the acoustic region as it is clear that these modes have maximum contribution in $\kappa_L$ of KMgSb (see SI). This rise in the $\gamma$ value indicates increasing anharmonicity in the vibration modes and leads to a significant drop in $\kappa_L$ of KMgSb (Fig.~\ref{fig:ltc}). Thus, the average and directional $\kappa_L$ decrease under the application of negative pressure. This reduction is purely due to the change in the bonding strength when the volume of the cell increases. The ambient pressure $\kappa_L$ obtained in our calculations is higher than the recently reported values \cite{2023_SHE}. The discrepancy is possibly due to the different techniques used for $\kappa_L$ calculations. Moreover, the methods used by Aixian et al. are also known to underestimate $\kappa_L$ in some cases \cite{Tadano_2014, Tadano_2015}. The reduced $\kappa_L$ helps improve the thermoelectric figure of merit (zT), subject to the condition that the power factor (PF = $S^{2}\sigma$) remains unchanged or increases. 
 
\begin{figure}[t]
		\centering\includegraphics[scale=0.65]{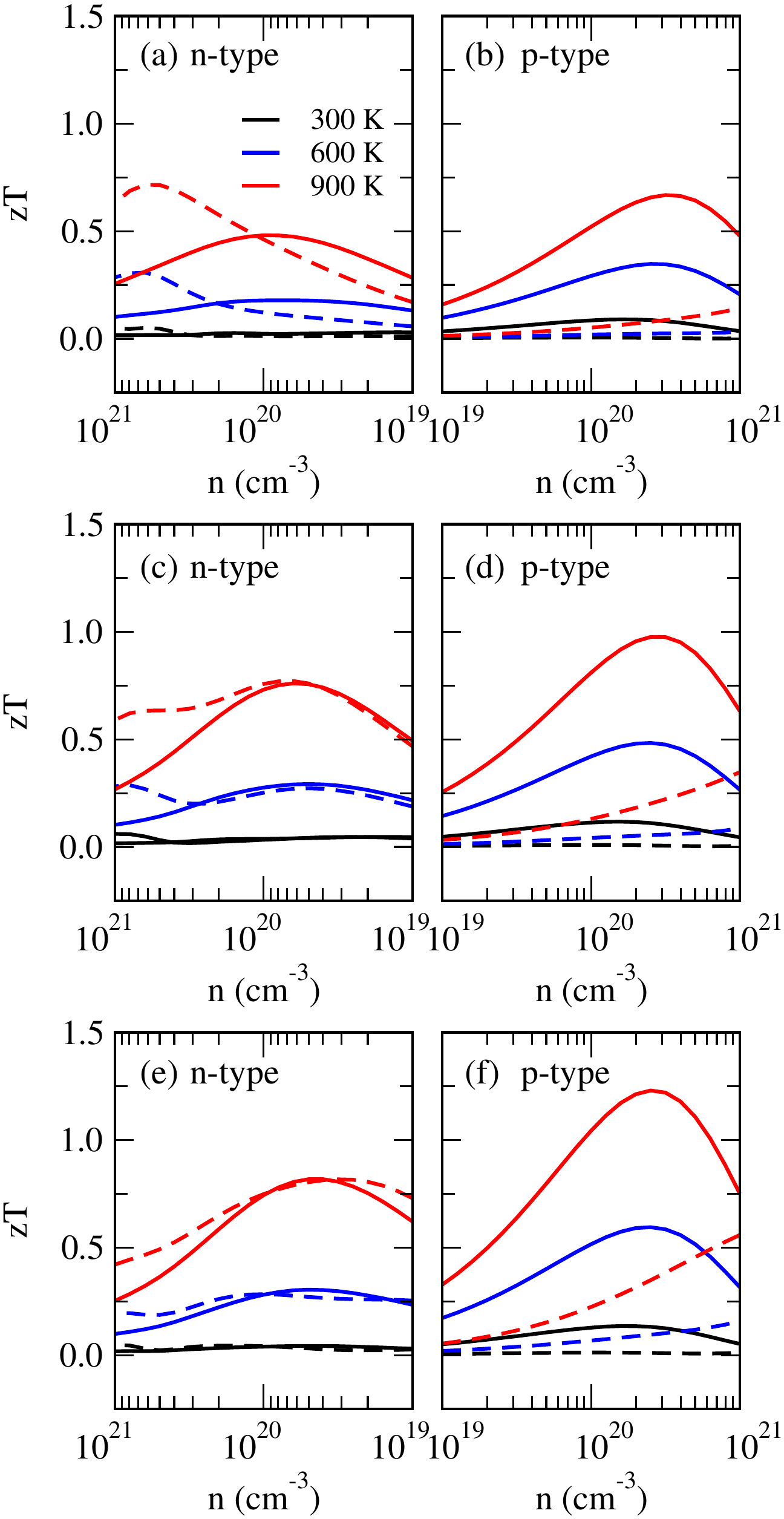}
		\caption{\label{fig:zT}The thermoelectric figure of merit of KMgSb at (a, b) 1.0~GPa, (c, d) ambient, and (e, f) -1.0~GPa pressure. The black, blue, and red curves represent zT at 300 K,  600 K, and 900 K temperatures, respectively. The solid and dotted curves show zT in the $a-b$ plane and along the $c$ direction. On $x-$axis, we have used $n$ for doping level.}
\end{figure}
We observed two increasing trends in the thermoelectric figure of merit (zT) values of KMgSb, one with increasing pressure (in $c$ direction and at $\sim6\times10^{20}~\mathrm{cm^{-3}}$ $n$-type doping) and the other with decreasing pressure (at $\sim3.9\times10^{19}~\mathrm{cm^{-3}}$ in $n$-type and at $\sim2.5\times10^{20}~\mathrm{cm^{-3}}$ in the $p$-type regions). Now, we will explain both trends one by one. The peak value of PF of KMgSb is found to increase with pressure in all directions and both doping regions. In contrast, little changes are observed in electronic and lattice thermal conductivity. Therefore, at $900$~K, the enhancement in zT value (see Fig.~\ref{fig:zT}) in the $n$-type doping is solely due to the increased PF. The electrical conductivity ($\sigma$) shows negligible variation with pressure; hence the changes in the PF are primarily due to the Seebeck coefficient (S), as shown in Supplementary information. The pressure changes the band dispersion in the $\Gamma\rightarrow Z$ direction of the conduction region, which is shown in Fig.~\ref{fig:band}. The bands inside the red circle (Fig.~\ref{fig:band}) are flatter at $1.0~$GPa pressure than the ambient and $-1.0~$GPa pressure, which results in increasing effective mass of electrons and Seebeck coefficient of KMgSb. This led to a sharp increase in the PF along the $c$ direction and in the $n$-type doping region, increasing the zT value. 

\begin{figure*}[ht]
	\centering\includegraphics[scale=0.8]{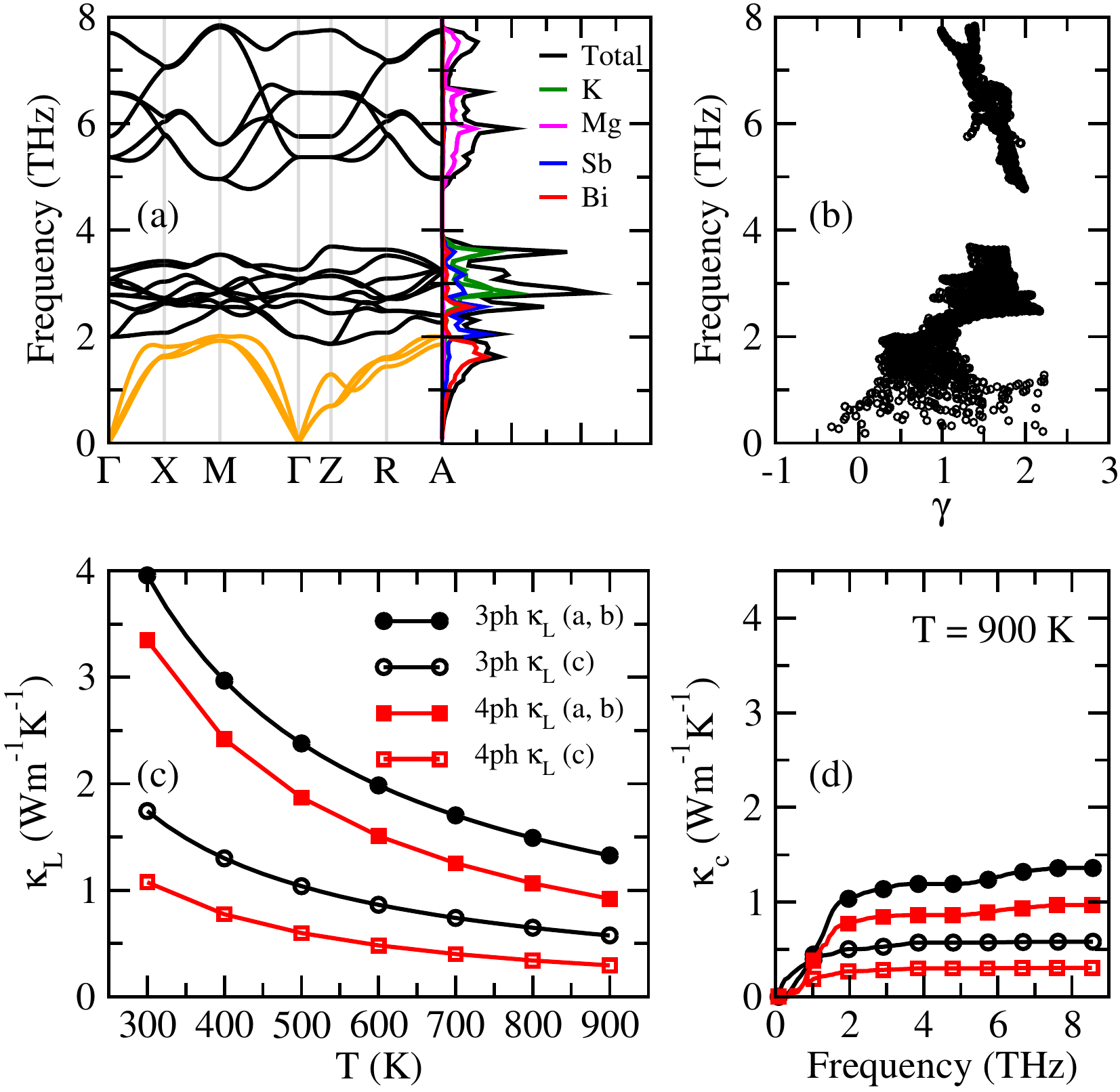}
	\caption{\label{fig:phonon}(a) Phonon band structure and density of states (PhDOS), (b) Gr\"uneisen parameter $\gamma$, (c) lattice thermal conductivity ($\kappa_{L}$), and (d) cumulative lattice thermal conductivity ($\kappa_{c}$) of KMgSb$_{0.5}$Bi$_{0.5}$ at $900$~K.}
\end{figure*}	
In the second case, the peak zT values are found to increase with decreasing pressure and both doping regions (see Fig.~\ref{fig:zT}). This increase in zT value is primarily due to the decreasing $\kappa_L$ (Fig.~\ref{fig:ltc}) because the PF does not increase as we go towards negative pressure. Therefore, the zT value can be increased by applying hydrostatic pressure. In experiments, positive pressure can be generated using a diamond anvil cell, whereas negative pressure generation seems extremely challenging. The negative pressure results in increased unit cell volume, which can be achieved by chemically doping larger atoms in place of smaller ones. Herein, we selected the Bi atom to replace Sb in KMgSb. Since Bi and Sb are iso-electronic, we expect that the alloying will only change the unit cell size and not the carrier concentration. This substitution is one way to create a negative pressure-like effect in experimental synthesis. As an interesting observation, we found that the $\kappa_L$ at -1.0~GPa matches closely (in \textit{a-b} plane) with the $\kappa_L$ of KMgSb$_{0.5}$Bi$_{0.5}$.  

The $n$-type and $p$-type zT values of KMgSb at $\sim8\times10^{19}~\mathrm{cm^{-3}}$ doping level are in close agreement. The ambient pressure zT value in the $n$-type and $p$-type region is $\sim0.75$ at $\sim8\times10^{19}~\mathrm{cm^{-3}}$, as shown in Fig.~\ref{fig:zT}~(b). This is an exciting feature of KMgSb, which is highly desirable in real-life thermoelectric device-based applications. Hence, KMgSb may be used as a $p$-type leg as well as an $n$-type leg of a thermoelectric device. Moreover, KMgSb is a $p$-type thermoelectric material above $10^{20}~\mathrm{cm^{-3}}$ carrier concentration. Most of the high-performance thermoelectric materials being reported are $n$-type, which makes KMgSb interesting as it may be used as a $p$-type leg in thermoelectric devices. 

\subsection{\label{subsec:Alloy}Effect of alloying }
Considering that the $\kappa_L$ of KMgSb decreases as we go towards negative pressure, it is worth exploring further. In a recent computational study, KMgBi was reported as an excellent thermoelectric material with a desirably high zT value ($>2.0$) \cite{KMgBi}. KMgSb being closer to KMgBi in size (than e.g. KMgAs), makes KMgSb an exciting candidate for experimentalists to replace Sb with Bi and fine-tune the thermal and electronic properties. We designed a composition by replacing 50 \% Sb atoms with Bi atoms. At -1.0~GPa, the $a$ and $c$ lattice constants of KMgSb are $4.853$ \AA\/ and $8.388$ \AA\/, respectively. The lattice constants of KMgSb$_{0.5}$Bi$_{0.5}$ (see Table \ref{tab:T1}) match closely with the lattice constant of KMgSb at -1.0~GPa. The difference between the $a$ and $c$ of KMgSb$_{0.5}$Bi$_{0.5}$ and KMgSb (at -1.0~GPa) is $0.019$~\AA\/ and $0.082$~\AA\/, respectively. Therefore, by doping 50 \% Bi in KMgSb, we have designed a composition of a size similar to KMgSb at -1.0~GPa. 	
	
The phonon band structure of KMgSb$_{0.5}$Bi$_{0.5}$ contains 18 bands, where the three lowest frequency bands (orange bands in Fig.~\ref{fig:phonon}~(a)) originating from $\Gamma$ point are the acoustic bands. All three acoustic modes, two transverse acoustic (TA) and one longitudinal acoustic (LA) are seen up to a frequency of $\sim$ 2~THz. These acoustic modes carry most of the heat in the crystal as observed in the cumulative lattice thermal conductivity ($\kappa_c$) shown in Fig.~\ref{fig:phonon}~(d). The phonon density of states (PhDOS) confirms that the acoustic modes are dominated by Sb atoms in KMgSb (see SI). The acoustic and optical branches of phonon bands of KMgSb become degenerate above 2.0~THz frequency. The doping of Bi in KMgSb directly reflects in acoustic branches of phonon bands. Bi doping lowers the maximum frequency of acoustic modes from $\sim$2.4~THz to $\sim$2.0~THz and separates the acoustic band from the optical bands. After Bi doping, the mode Gr\"uneisen parameter (Fig.~\ref{fig:phonon}~(b)) shows increased anharmonicity in the low-frequency modes very similar to KMgSb at -1.0~GPa, and these modes contribute more than 75 \% ($a-b$ plane) and 85 \% ($c$ direction) in $\kappa_L$ as shown in Fig.~\ref{fig:phonon}~(d). Bi doping does two interesting things; (a) increase the size of the cell and (b) causes strain in the lattice. Therefore, replacing Sb with Bi compensates for the negative pressure and also increases the anharmonicity in phonon modes due to the strain in lattice from the substitutional defects. These factors lead to a decrease in the $\kappa_L$ after doping Bi in KMgSb as shown in Fig.~\ref{fig:phonon}~(c).
	
\begin{figure}[t]
	\centering\includegraphics[scale=0.3]{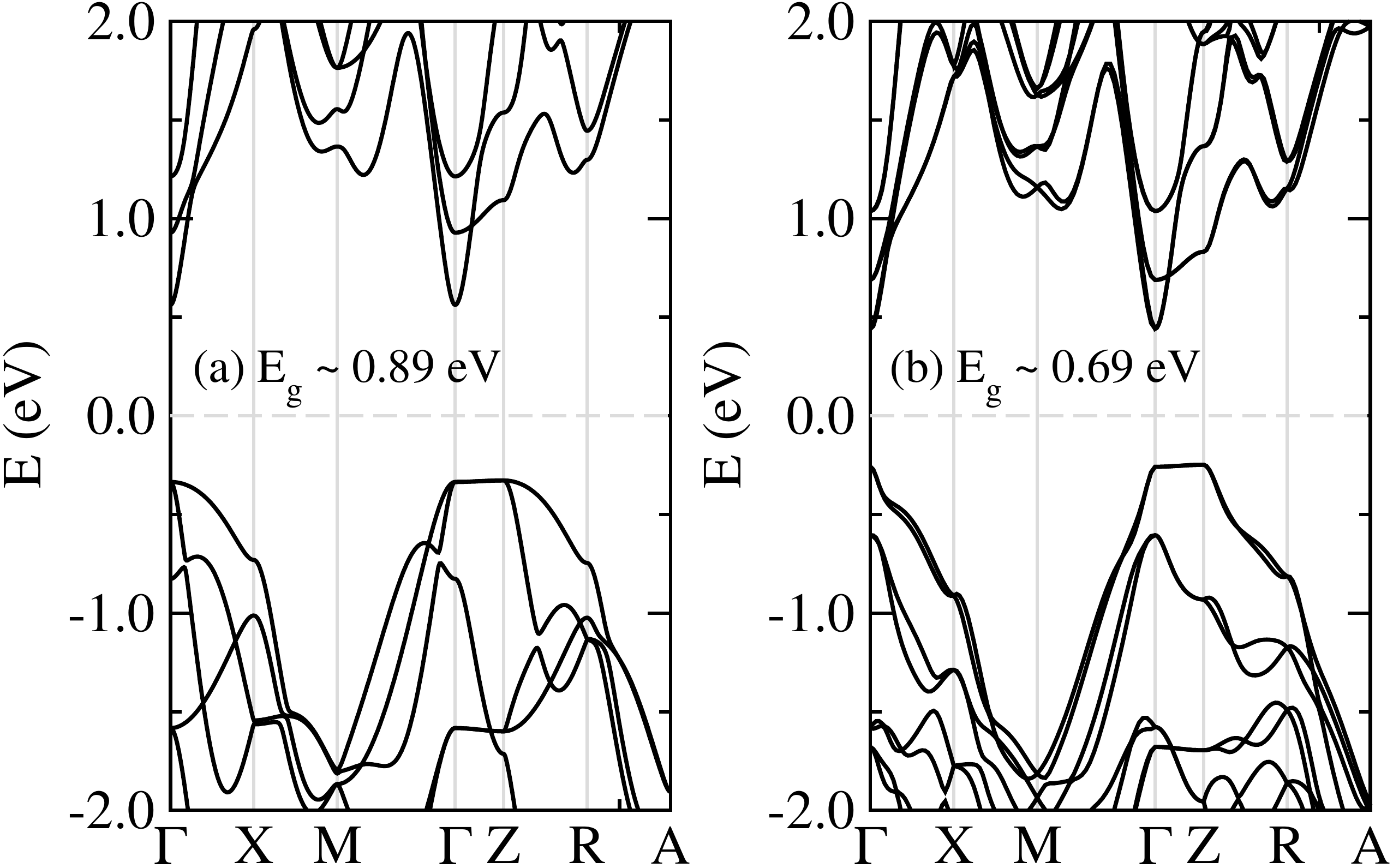}
	\caption{\label{fig:BS}The electronic band structure of KMgSb$_{0.5}$Bi$_{0.5}$ without (a) and with (b) spin-orbit coupling.}
	\end{figure}
	
\begin{figure}[h]
	\centering\includegraphics[scale=0.65]{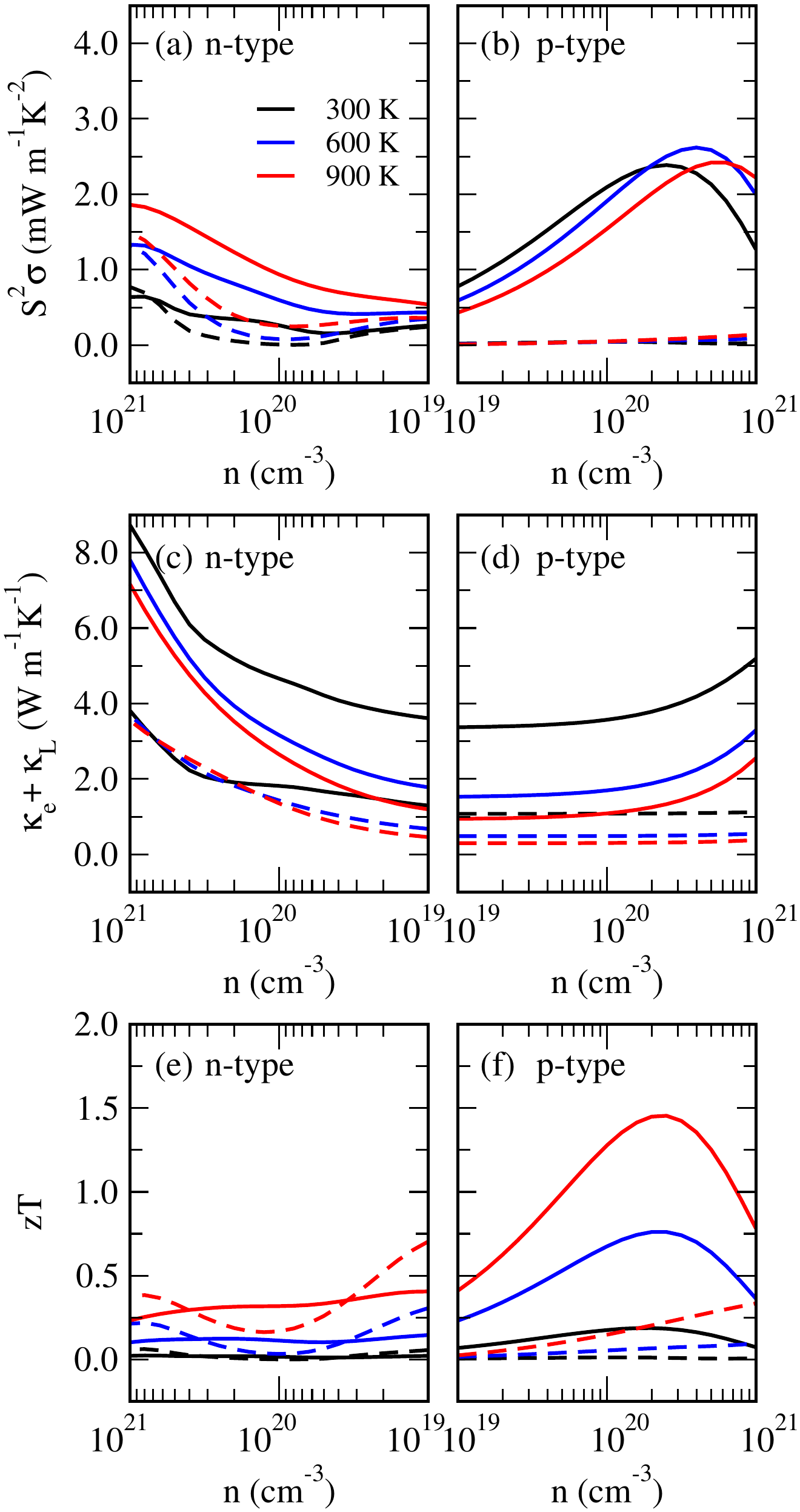}
	\caption{\label{fig:50trans}(a, b) The power factor (PF), (c, d) thermal conductivity ($\kappa_e$ + $\kappa_L$), and (e, f) thermoelectric figure of merit (zT) of KMgSb$_{0.5}$Bi$_{0.5}$. The black, blue, and red curves represent zT at 300 K,  600 K, and 900 K temperatures, respectively. The solid and dotted curves show zT in the $a-b$ plane and along the $c$ direction. On $x-$axis, we have used $n$ for doping level.}
\end{figure}
	
\begin{figure*}[t]
		\centering\includegraphics[scale=0.48]{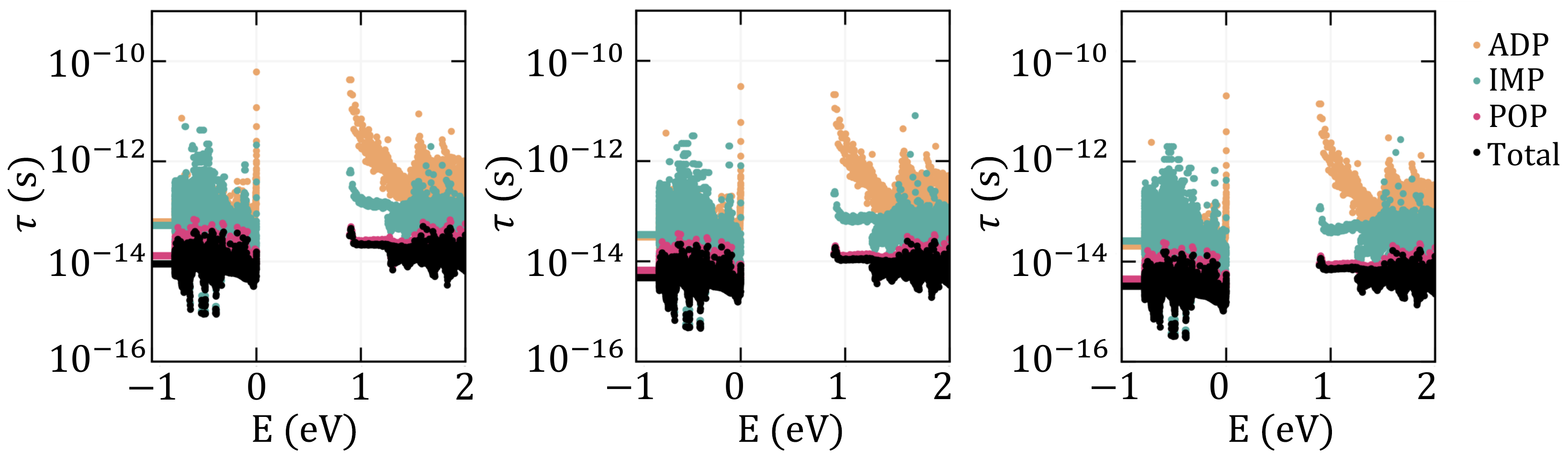}
		\caption{\label{fig:tau}The scattering mechanism dependent electronic relaxation time ($\tau$) in KMgSb$_{0.5}$Bi$_{0.5}$, calculated at $300$~K, $600$~K, and $900$~K for $1.26\times10^{20}~cm^{-3}$ $p$-type carrier concentration.}
\end{figure*}
	
In Fig.~\ref{fig:BS}, the band structure of KMgSb$_{0.5}$Bi$_{0.5}$ is shown, where flat (in $\Gamma\rightarrow Z$ direction and valance region) as well as largely dispersive (in conduction region) bands are seen. The band gap of KMgSb$_{0.5}$Bi$_{0.5}$ is $\sim$0.89~eV without spin-orbit coupling (SOC), which drops to $\sim$0.69~eV with SOC. Bi doping has significantly reduced the band gap. Such tunability of band gap (with doping) is highly desirable while designing materials for thermoelectric applications. In Fig.~\ref{fig:50trans}, we have presented the power factor (PF), total thermal conductivity ($\kappa_e+\kappa_L$), and zT of KMgSb$_{0.5}$Bi$_{0.5}$. The zT value of KMgSb and KMgSb$_{0.5}$Bi$_{0.5}$ increases with temperature. At $900$~K, the peak zT value of KMgSb$_{0.5}$Bi$_{0.5}$ is $\sim1.45$, which is obtained in the $p$-type region as shown in Fig.~\ref{fig:50trans}~(c). The $\kappa_e$ of KMgSb and KMgSb$_{0.5}$Bi$_{0.5}$ at $900$~K are closely matching (see Supplementary information) at the doping level of peak zT value, whereas a significant increase in PF ($S^{2}\sigma$) is observed after Bi doping. If we compare the peak zT value of KMgSb and KMgSb$_{0.5}$Bi$_{0.5}$ in the $p$-type region, an increase of $\sim49~\%$ is obtained after alloy engineering. This increase in zT value is due to the increased PF and reduced $\kappa_L$ after Bi doping in KMgSb. As discussed above, KMgSb remains $p$-type thermoelectric material under hydrostatic pressure. Bi doping enhances zT value, and KMgSb$_{0.5}$Bi$_{0.5}$ becomes a potential high-performance $p$-type thermoelectric material, which is highly desirable for device engineering. 
	
As discussed above, substitutional doping creates defects that increase phonon scattering and reduces $\kappa_L$, which is desirable for thermoelectric applications. We also investigated the thermoelectric transport of As doped KMgSb but observed no significant increase or decrease in the peak zT value, even though the $\kappa_L$ improves after As doping. The zT value of KMgSb$_{0.5}$As$_{0.5}$ remains close to the zT value of KMgSb at ambient pressure. The electronic and ionic properties of KMgSb$_{0.5}$As$_{0.5}$ are given in Supplementary information. 
	
One important aspect of thermoelectric transport is the electronic relaxation time, which is a complex phenomenon. As discussed in computational details, the CRTA was considered to be the most convenient and helpful choice in electronic transport calculations, but it comes with a cost as under CRTA, the Seebeck and Hall coefficients become independent of scattering rates \cite{CRTA}. Moreover, a constant value of $\tau$ cannot account for all the scattering processes involved in electronic transport. To get an insight, we have shown the scattering mechanism dependent relaxation time ($\tau$) in Fig.~\ref{fig:tau}. The overall electronic relaxation time is found to lie between $\sim10^{-13}~$s and $\sim10^{-16}~$s, which arises primarily from the POP with a small contribution from ADP and IMP scattering mechanisms. 
	
The results discussed above explain the impact of pressure and alloy engineering on thermal and electronic transport. The thermoelectric performance shows a huge improvement after Bi doping. The spin-orbit coupling (SOC) is also an important factor in transport properties. It requires a lot of resources to include SOC in all DFT, hybrid DFT, and thermal and electronic transport calculations. The Seebeck coefficient ($S$), electronic conductivity ($\sigma$), and electronic component of thermal conductivity ($\kappa_e$) of KMgSb and KMgSb$_{0.5}$Bi$_{0.5}$ with and without SOC is given in Supplementary information. The calculation of $\kappa_L$ (including fourth-order IFCs) with SOC is extremely challenging and not included in this study. We observed the SOC has a small effect on $S$, $\sigma$, and $\kappa_e$. 
	
\section{\label{sec:conc}Summary and Conclusions}
In this work, we started with the investigation of the crystal and electronic structure of the KMgX family (X =  P, As, and Sb). All compounds of the KMgX family have van der Waals interaction along the $c$ direction. The van der Waals interaction is treated with optB86b-vdW with PBE functional, which we found to give lattice constants closely matching with the experimental values. The effect of hydrostatic pressure and alloy engineering on the electronic and thermal transport properties of KMgSb is investigated in this work.
	
It is encouraging to note that the $\kappa_L$ of KMgSb decreases under negative pressure, implying the application of negative pressure would be helpful in improving the thermoelectric performance. We observed that the zT value shows two increasing trends with the applied pressure. The positive pressure improves the zT value in the $n$-type region and along the $c$ direction. A significant increase in the zT values in the $n$-type, as well as the $p$-type region, is observed with the negative pressure. Interestingly, the $n$-type and $p$-type zT values of KMgSb at ambient and $-1.0~$GPa pressure are in close agreement at $\sim8\times10^{19}~\mathrm{cm^{-3}}$ carrier concentration. This is an important feature that is desirable and can be utilized in thermoelectric device engineering.
	
Pressure plays an important role. Negative pressure, which is very challenging experimentally, increases the volume of the cell. In order to bring in the negative pressure-like effect, we substituted 50 \% Sb with Bi atoms. The alloy engineering does improve thermoelectric performance dramatically. Bi doping decreases the $\kappa_L$ but increases the power factor, and the overall effect leads to a $\sim49~\%$ increase in the zT value in the $a-b$ plane and $p$-type doping region. The $\kappa_L$ (in $a-b$ plane) of KMgSb$_{0.5}$Bi$_{0.5}$ matches well with the $\kappa_L$ of KMgSb at $-1.0~$GPa pressure. Therefore, the negative pressure-like effect can possibly be obtained via chemical doping, at least as far as phonon transport is concerned. To conclude, hydrostatic pressure and alloy engineering help in improving the thermoelectric performance of quasi-2D material KMgSb. Also, KMgSb can be used as an $n$-type as well as $p$-type thermoelectric material. 

The supplementary data is available here \cite{supp}.
	
\begin{acknowledgments}
This work used the Supercomputing facility of IIT Roorkee established under National Supercomputing Mission (NSM), Government of India and supported by Centre for Development of Advanced Computing (CDAC), Pune. We have also used other computational facilities provided by Institute Computer Center (ICC), IIT Roorkee. VC wish to acknowledge the financial support received from Ministry of Education, Government of India.
\end{acknowledgments}
	

\begin{thebibliography}{56}%
	\makeatletter
	\providecommand \@ifxundefined [1]{%
		\@ifx{#1\undefined}
	}%
	\providecommand \@ifnum [1]{%
		\ifnum #1\expandafter \@firstoftwo
		\else \expandafter \@secondoftwo
		\fi
	}%
	\providecommand \@ifx [1]{%
		\ifx #1\expandafter \@firstoftwo
		\else \expandafter \@secondoftwo
		\fi
	}%
	\providecommand \natexlab [1]{#1}%
	\providecommand \enquote  [1]{``#1''}%
	\providecommand \bibnamefont  [1]{#1}%
	\providecommand \bibfnamefont [1]{#1}%
	\providecommand \citenamefont [1]{#1}%
	\providecommand \href@noop [0]{\@secondoftwo}%
	\providecommand \href [0]{\begingroup \@sanitize@url \@href}%
	\providecommand \@href[1]{\@@startlink{#1}\@@href}%
	\providecommand \@@href[1]{\endgroup#1\@@endlink}%
	\providecommand \@sanitize@url [0]{\catcode `\\12\catcode `\$12\catcode
		`\&12\catcode `\#12\catcode `\^12\catcode `\_12\catcode `\%12\relax}%
	\providecommand \@@startlink[1]{}%
	\providecommand \@@endlink[0]{}%
	\providecommand \url  [0]{\begingroup\@sanitize@url \@url }%
	\providecommand \@url [1]{\endgroup\@href {#1}{\urlprefix }}%
	\providecommand \urlprefix  [0]{URL }%
	\providecommand \Eprint [0]{\href }%
	\providecommand \doibase [0]{https://doi.org/}%
	\providecommand \selectlanguage [0]{\@gobble}%
	\providecommand \bibinfo  [0]{\@secondoftwo}%
	\providecommand \bibfield  [0]{\@secondoftwo}%
	\providecommand \translation [1]{[#1]}%
	\providecommand \BibitemOpen [0]{}%
	\providecommand \bibitemStop [0]{}%
	\providecommand \bibitemNoStop [0]{.\EOS\space}%
	\providecommand \EOS [0]{\spacefactor3000\relax}%
	\providecommand \BibitemShut  [1]{\csname bibitem#1\endcsname}%
	\let\auto@bib@innerbib\@empty
	\bibitem [{\citenamefont {Snyder}\ and\ \citenamefont {Toberer}(2008)}]{Rev1}%
	\BibitemOpen
	\bibfield  {author} {\bibinfo {author} {\bibfnamefont {G.~J.}\ \bibnamefont
			{Snyder}}\ and\ \bibinfo {author} {\bibfnamefont {E.~S.}\ \bibnamefont
			{Toberer}},\ }\bibfield  {title} {\bibinfo {title} {Complex thermoelectric
			materials},\ }\href {https://doi.org/10.1038/nmat2090} {\bibfield  {journal}
		{\bibinfo  {journal} {Nat. Mater.}\ }\textbf {\bibinfo {volume} {7}},\
		\bibinfo {pages} {105} (\bibinfo {year} {2008})}\BibitemShut {NoStop}%
	\bibitem [{\citenamefont {Wei}\ \emph {et~al.}(2020)\citenamefont {Wei},
		\citenamefont {Yang}, \citenamefont {Ma}, \citenamefont {Song}, \citenamefont
		{Zhang}, \citenamefont {Ma}, \citenamefont {Yang},\ and\ \citenamefont
		{Wang}}]{Rev2}%
	\BibitemOpen
	\bibfield  {author} {\bibinfo {author} {\bibfnamefont {J.}~\bibnamefont
			{Wei}}, \bibinfo {author} {\bibfnamefont {L.}~\bibnamefont {Yang}}, \bibinfo
		{author} {\bibfnamefont {Z.}~\bibnamefont {Ma}}, \bibinfo {author}
		{\bibfnamefont {P.}~\bibnamefont {Song}}, \bibinfo {author} {\bibfnamefont
			{M.}~\bibnamefont {Zhang}}, \bibinfo {author} {\bibfnamefont
			{J.}~\bibnamefont {Ma}}, \bibinfo {author} {\bibfnamefont {F.}~\bibnamefont
			{Yang}},\ and\ \bibinfo {author} {\bibfnamefont {X.}~\bibnamefont {Wang}},\
	}\bibfield  {title} {\bibinfo {title} {Review of current high-$\mathrm{ZT}$
			thermoelectric materials},\ }\href
	{https://doi.org/10.1007/S10853-020-04949-0} {\bibfield  {journal} {\bibinfo
			{journal} {J. Mater. Sci.}\ }\textbf {\bibinfo {volume} {55}},\ \bibinfo
		{pages} {12642} (\bibinfo {year} {2020})}\BibitemShut {NoStop}%
	\bibitem [{\citenamefont {Kim}\ \emph {et~al.}(2015)\citenamefont {Kim},
		\citenamefont {Lee}, \citenamefont {Mun}, \citenamefont {Kim}, \citenamefont
		{Hwang}, \citenamefont {Roh}, \citenamefont {Yang}, \citenamefont {Shin},
		\citenamefont {Li}, \citenamefont {Lee}, \citenamefont {Snyder},\ and\
		\citenamefont {Kim}}]{ref1}%
	\BibitemOpen
	\bibfield  {author} {\bibinfo {author} {\bibfnamefont {S.~I.}\ \bibnamefont
			{Kim}}, \bibinfo {author} {\bibfnamefont {K.~H.}\ \bibnamefont {Lee}},
		\bibinfo {author} {\bibfnamefont {H.~A.}\ \bibnamefont {Mun}}, \bibinfo
		{author} {\bibfnamefont {H.~S.}\ \bibnamefont {Kim}}, \bibinfo {author}
		{\bibfnamefont {S.~W.}\ \bibnamefont {Hwang}}, \bibinfo {author}
		{\bibfnamefont {J.~W.}\ \bibnamefont {Roh}}, \bibinfo {author} {\bibfnamefont
			{D.~J.}\ \bibnamefont {Yang}}, \bibinfo {author} {\bibfnamefont {W.~H.}\
			\bibnamefont {Shin}}, \bibinfo {author} {\bibfnamefont {X.~S.}\ \bibnamefont
			{Li}}, \bibinfo {author} {\bibfnamefont {Y.~H.}\ \bibnamefont {Lee}},
		\bibinfo {author} {\bibfnamefont {G.~J.}\ \bibnamefont {Snyder}},\ and\
		\bibinfo {author} {\bibfnamefont {S.~W.}\ \bibnamefont {Kim}},\ }\bibfield
	{title} {\bibinfo {title} {Dense dislocation arrays embedded in grain
			boundaries for high-performance bulk thermoelectrics},\ }\href
	{https://doi.org/10.1126/science.aaa4166} {\bibfield  {journal} {\bibinfo
			{journal} {Science}\ }\textbf {\bibinfo {volume} {348}},\ \bibinfo {pages}
		{109} (\bibinfo {year} {2015})}\BibitemShut {NoStop}%
	\bibitem [{\citenamefont {Poudel}\ \emph {et~al.}(2008)\citenamefont {Poudel},
		\citenamefont {Hao}, \citenamefont {Ma}, \citenamefont {Lan}, \citenamefont
		{Minnich}, \citenamefont {Yu}, \citenamefont {Yan}, \citenamefont {Wang},
		\citenamefont {Muto}, \citenamefont {Vashaee}, \citenamefont {Chen},
		\citenamefont {Liu}, \citenamefont {Dresselhaus}, \citenamefont {Chen},\ and\
		\citenamefont {Ren}}]{ref2}%
	\BibitemOpen
	\bibfield  {author} {\bibinfo {author} {\bibfnamefont {B.}~\bibnamefont
			{Poudel}}, \bibinfo {author} {\bibfnamefont {Q.}~\bibnamefont {Hao}},
		\bibinfo {author} {\bibfnamefont {Y.}~\bibnamefont {Ma}}, \bibinfo {author}
		{\bibfnamefont {Y.}~\bibnamefont {Lan}}, \bibinfo {author} {\bibfnamefont
			{A.}~\bibnamefont {Minnich}}, \bibinfo {author} {\bibfnamefont
			{B.}~\bibnamefont {Yu}}, \bibinfo {author} {\bibfnamefont {X.}~\bibnamefont
			{Yan}}, \bibinfo {author} {\bibfnamefont {D.}~\bibnamefont {Wang}}, \bibinfo
		{author} {\bibfnamefont {A.}~\bibnamefont {Muto}}, \bibinfo {author}
		{\bibfnamefont {D.}~\bibnamefont {Vashaee}}, \bibinfo {author} {\bibfnamefont
			{X.}~\bibnamefont {Chen}}, \bibinfo {author} {\bibfnamefont {J.}~\bibnamefont
			{Liu}}, \bibinfo {author} {\bibfnamefont {M.~S.}\ \bibnamefont
			{Dresselhaus}}, \bibinfo {author} {\bibfnamefont {G.}~\bibnamefont {Chen}},\
		and\ \bibinfo {author} {\bibfnamefont {Z.}~\bibnamefont {Ren}},\ }\bibfield
	{title} {\bibinfo {title} {High-thermoelectric performance of nanostructured
			$\mathrm{B}$ismuth $\mathrm{A}$ntimony $\mathrm{T}$elluride bulk alloys},\
	}\href {https://doi.org/10.1126/science.1156446} {\bibfield  {journal}
		{\bibinfo  {journal} {Science}\ }\textbf {\bibinfo {volume} {320}},\ \bibinfo
		{pages} {634} (\bibinfo {year} {2008})}\BibitemShut {NoStop}%
	\bibitem [{\citenamefont {Hu}\ \emph {et~al.}()\citenamefont {Hu},
		\citenamefont {Wu}, \citenamefont {Zhu}, \citenamefont {Fu}, \citenamefont
		{He}, \citenamefont {Ying},\ and\ \citenamefont {Zhao}}]{ref3}%
	\BibitemOpen
	\bibfield  {author} {\bibinfo {author} {\bibfnamefont {L.}~\bibnamefont
			{Hu}}, \bibinfo {author} {\bibfnamefont {H.}~\bibnamefont {Wu}}, \bibinfo
		{author} {\bibfnamefont {T.}~\bibnamefont {Zhu}}, \bibinfo {author}
		{\bibfnamefont {C.}~\bibnamefont {Fu}}, \bibinfo {author} {\bibfnamefont
			{J.}~\bibnamefont {He}}, \bibinfo {author} {\bibfnamefont {P.}~\bibnamefont
			{Ying}},\ and\ \bibinfo {author} {\bibfnamefont {X.}~\bibnamefont {Zhao}},\
	}\bibfield  {title} {\bibinfo {title} {Tuning multiscale microstructures to
			enhance thermoelectric performance of n-type
			$\mathrm{B}$ismuth-$\mathrm{T}$elluride-based solid solutions},\ }\href
	{https://doi.org/https://doi.org/10.1002/aenm.201500411} {\bibfield
		{journal} {\bibinfo  {journal} {Adv. Energy Mater.}\ }\textbf {\bibinfo
			{volume} {5}},\ \bibinfo {pages} {1500411}}\BibitemShut {NoStop}%
	\bibitem [{\citenamefont {Zhao}\ \emph {et~al.}(2014)\citenamefont {Zhao},
		\citenamefont {Sui}, \citenamefont {Tang}, \citenamefont {Lan}, \citenamefont
		{Jie}, \citenamefont {Kraemer}, \citenamefont {McEnaney}, \citenamefont
		{Guloy}, \citenamefont {Chen},\ and\ \citenamefont {Ren}}]{ref4}%
	\BibitemOpen
	\bibfield  {author} {\bibinfo {author} {\bibfnamefont {H.}~\bibnamefont
			{Zhao}}, \bibinfo {author} {\bibfnamefont {J.}~\bibnamefont {Sui}}, \bibinfo
		{author} {\bibfnamefont {Z.}~\bibnamefont {Tang}}, \bibinfo {author}
		{\bibfnamefont {Y.}~\bibnamefont {Lan}}, \bibinfo {author} {\bibfnamefont
			{Q.}~\bibnamefont {Jie}}, \bibinfo {author} {\bibfnamefont {D.}~\bibnamefont
			{Kraemer}}, \bibinfo {author} {\bibfnamefont {K.}~\bibnamefont {McEnaney}},
		\bibinfo {author} {\bibfnamefont {A.}~\bibnamefont {Guloy}}, \bibinfo
		{author} {\bibfnamefont {G.}~\bibnamefont {Chen}},\ and\ \bibinfo {author}
		{\bibfnamefont {Z.}~\bibnamefont {Ren}},\ }\bibfield  {title} {\bibinfo
		{title} {High thermoelectric performance of {MgAgSb}-based materials},\
	}\href {https://doi.org/https://doi.org/10.1016/j.nanoen.2014.04.012}
	{\bibfield  {journal} {\bibinfo  {journal} {Nano Energy}\ }\textbf {\bibinfo
			{volume} {7}},\ \bibinfo {pages} {97} (\bibinfo {year} {2014})}\BibitemShut
	{NoStop}%
	\bibitem [{\citenamefont {Zhao}\ \emph {et~al.}(2012)\citenamefont {Zhao},
		\citenamefont {He}, \citenamefont {Wu}, \citenamefont {Hogan}, \citenamefont
		{Zhou}, \citenamefont {Uher}, \citenamefont {Dravid},\ and\ \citenamefont
		{Kanatzidis}}]{1PbTe1}%
	\BibitemOpen
	\bibfield  {author} {\bibinfo {author} {\bibfnamefont {L.-D.}\ \bibnamefont
			{Zhao}}, \bibinfo {author} {\bibfnamefont {J.}~\bibnamefont {He}}, \bibinfo
		{author} {\bibfnamefont {C.-I.}\ \bibnamefont {Wu}}, \bibinfo {author}
		{\bibfnamefont {T.~P.}\ \bibnamefont {Hogan}}, \bibinfo {author}
		{\bibfnamefont {X.}~\bibnamefont {Zhou}}, \bibinfo {author} {\bibfnamefont
			{C.}~\bibnamefont {Uher}}, \bibinfo {author} {\bibfnamefont {V.~P.}\
			\bibnamefont {Dravid}},\ and\ \bibinfo {author} {\bibfnamefont {M.~G.}\
			\bibnamefont {Kanatzidis}},\ }\bibfield  {title} {\bibinfo {title}
		{Thermoelectrics with earth abundant elements: High performance p-type {PbS}
			nanostructured with {SrS} and {CaS}},\ }\href
	{https://doi.org/10.1021/ja301772w} {\bibfield  {journal} {\bibinfo
			{journal} {J. Am. Chem. Soc.}\ }\textbf {\bibinfo {volume} {134}},\ \bibinfo
		{pages} {7902} (\bibinfo {year} {2012})}\BibitemShut {NoStop}%
	\bibitem [{\citenamefont {Zhao}\ \emph
		{et~al.}(2016{\natexlab{a}})\citenamefont {Zhao}, \citenamefont {Tan},
		\citenamefont {Hao}, \citenamefont {He}, \citenamefont {Pei}, \citenamefont
		{Chi}, \citenamefont {Wang}, \citenamefont {Gong}, \citenamefont {Xu},
		\citenamefont {Dravid}, \citenamefont {Uher}, \citenamefont {Snyder},
		\citenamefont {Wolverton},\ and\ \citenamefont {Kanatzidis}}]{ref6}%
	\BibitemOpen
	\bibfield  {author} {\bibinfo {author} {\bibfnamefont {L.-D.}\ \bibnamefont
			{Zhao}}, \bibinfo {author} {\bibfnamefont {G.}~\bibnamefont {Tan}}, \bibinfo
		{author} {\bibfnamefont {S.}~\bibnamefont {Hao}}, \bibinfo {author}
		{\bibfnamefont {J.}~\bibnamefont {He}}, \bibinfo {author} {\bibfnamefont
			{Y.}~\bibnamefont {Pei}}, \bibinfo {author} {\bibfnamefont {H.}~\bibnamefont
			{Chi}}, \bibinfo {author} {\bibfnamefont {H.}~\bibnamefont {Wang}}, \bibinfo
		{author} {\bibfnamefont {S.}~\bibnamefont {Gong}}, \bibinfo {author}
		{\bibfnamefont {H.}~\bibnamefont {Xu}}, \bibinfo {author} {\bibfnamefont
			{V.~P.}\ \bibnamefont {Dravid}}, \bibinfo {author} {\bibfnamefont
			{C.}~\bibnamefont {Uher}}, \bibinfo {author} {\bibfnamefont {G.~J.}\
			\bibnamefont {Snyder}}, \bibinfo {author} {\bibfnamefont {C.}~\bibnamefont
			{Wolverton}},\ and\ \bibinfo {author} {\bibfnamefont {M.~G.}\ \bibnamefont
			{Kanatzidis}},\ }\bibfield  {title} {\bibinfo {title} {Ultrahigh power factor
			and thermoelectric performance in hole-doped single-crystal {SnSe}},\ }\href
	{https://doi.org/10.1126/science.aad3749} {\bibfield  {journal} {\bibinfo
			{journal} {Science}\ }\textbf {\bibinfo {volume} {351}},\ \bibinfo {pages}
		{141} (\bibinfo {year} {2016}{\natexlab{a}})}\BibitemShut {NoStop}%
	\bibitem [{\citenamefont {Zhou}\ and\ \citenamefont {Zhao}()}]{ref7}%
	\BibitemOpen
	\bibfield  {author} {\bibinfo {author} {\bibfnamefont {Y.}~\bibnamefont
			{Zhou}}\ and\ \bibinfo {author} {\bibfnamefont {L.-D.}\ \bibnamefont
			{Zhao}},\ }\bibfield  {title} {\bibinfo {title} {Promising thermoelectric
			bulk materials with {2D} structures},\ }\href
	{https://doi.org/10.1002/adma.201702676} {\bibfield  {journal} {\bibinfo
			{journal} {Adv. Mater.}\ }\textbf {\bibinfo {volume} {29}},\ \bibinfo {pages}
		{1702676}}\BibitemShut {NoStop}%
	\bibitem [{\citenamefont {Chen}\ \emph {et~al.}(2018)\citenamefont {Chen},
		\citenamefont {Shi}, \citenamefont {Zhao},\ and\ \citenamefont
		{Zou}}]{2SnSe1}%
	\BibitemOpen
	\bibfield  {author} {\bibinfo {author} {\bibfnamefont {Z.-G.}\ \bibnamefont
			{Chen}}, \bibinfo {author} {\bibfnamefont {X.}~\bibnamefont {Shi}}, \bibinfo
		{author} {\bibfnamefont {L.-D.}\ \bibnamefont {Zhao}},\ and\ \bibinfo
		{author} {\bibfnamefont {J.}~\bibnamefont {Zou}},\ }\bibfield  {title}
	{\bibinfo {title} {High-performance {SnSe} thermoelectric materials: Progress
			and future challenge},\ }\href
	{https://doi.org/10.1016/j.pmatsci.2018.04.005} {\bibfield  {journal}
		{\bibinfo  {journal} {Prog. Mater. Sci.}\ }\textbf {\bibinfo {volume} {97}},\
		\bibinfo {pages} {283} (\bibinfo {year} {2018})}\BibitemShut {NoStop}%
	\bibitem [{\citenamefont {Zhao}\ \emph
		{et~al.}(2016{\natexlab{b}})\citenamefont {Zhao}, \citenamefont {Chang},
		\citenamefont {Tan},\ and\ \citenamefont {Kanatzidis}}]{3SnSe2}%
	\BibitemOpen
	\bibfield  {author} {\bibinfo {author} {\bibfnamefont {L.-D.}\ \bibnamefont
			{Zhao}}, \bibinfo {author} {\bibfnamefont {C.}~\bibnamefont {Chang}},
		\bibinfo {author} {\bibfnamefont {G.}~\bibnamefont {Tan}},\ and\ \bibinfo
		{author} {\bibfnamefont {M.~G.}\ \bibnamefont {Kanatzidis}},\ }\bibfield
	{title} {\bibinfo {title} {{SnSe}: a remarkable new thermoelectric
			material},\ }\href {https://doi.org/10.1039/C6EE01755J} {\bibfield  {journal}
		{\bibinfo  {journal} {Energy Environ. Sci.}\ }\textbf {\bibinfo {volume}
			{9}},\ \bibinfo {pages} {3044} (\bibinfo {year}
		{2016}{\natexlab{b}})}\BibitemShut {NoStop}%
	\bibitem [{\citenamefont {Zhao}\ \emph
		{et~al.}(2016{\natexlab{c}})\citenamefont {Zhao}, \citenamefont {Zhang},
		\citenamefont {Wu}, \citenamefont {Tan}, \citenamefont {Pei}, \citenamefont
		{Xiao}, \citenamefont {Chang}, \citenamefont {Wu}, \citenamefont {Chi},
		\citenamefont {Zheng}, \citenamefont {Gong}, \citenamefont {Uher},
		\citenamefont {He},\ and\ \citenamefont {Kanatzidis}}]{4SnTe}%
	\BibitemOpen
	\bibfield  {author} {\bibinfo {author} {\bibfnamefont {L.-D.}\ \bibnamefont
			{Zhao}}, \bibinfo {author} {\bibfnamefont {X.}~\bibnamefont {Zhang}},
		\bibinfo {author} {\bibfnamefont {H.}~\bibnamefont {Wu}}, \bibinfo {author}
		{\bibfnamefont {G.}~\bibnamefont {Tan}}, \bibinfo {author} {\bibfnamefont
			{Y.}~\bibnamefont {Pei}}, \bibinfo {author} {\bibfnamefont {Y.}~\bibnamefont
			{Xiao}}, \bibinfo {author} {\bibfnamefont {C.}~\bibnamefont {Chang}},
		\bibinfo {author} {\bibfnamefont {D.}~\bibnamefont {Wu}}, \bibinfo {author}
		{\bibfnamefont {H.}~\bibnamefont {Chi}}, \bibinfo {author} {\bibfnamefont
			{L.}~\bibnamefont {Zheng}}, \bibinfo {author} {\bibfnamefont
			{S.}~\bibnamefont {Gong}}, \bibinfo {author} {\bibfnamefont {C.}~\bibnamefont
			{Uher}}, \bibinfo {author} {\bibfnamefont {J.}~\bibnamefont {He}},\ and\
		\bibinfo {author} {\bibfnamefont {M.~G.}\ \bibnamefont {Kanatzidis}},\
	}\bibfield  {title} {\bibinfo {title} {Enhanced thermoelectric properties in
			the counter-doped {SnTe} system with strained endotaxial {SrTe}},\ }\href
	{https://doi.org/10.1021/jacs.5b13276} {\bibfield  {journal} {\bibinfo
			{journal} {J. Am. Chem. Soc.}\ }\textbf {\bibinfo {volume} {138}},\ \bibinfo
		{pages} {2366} (\bibinfo {year} {2016}{\natexlab{c}})}\BibitemShut {NoStop}%
	\bibitem [{\citenamefont {Wu}\ \emph {et~al.}(2014)\citenamefont {Wu},
		\citenamefont {Zhao}, \citenamefont {Hao}, \citenamefont {Jiang},
		\citenamefont {Zheng}, \citenamefont {Doak}, \citenamefont {Wu},
		\citenamefont {Chi}, \citenamefont {Gelbstein}, \citenamefont {Uher},
		\citenamefont {Wolverton}, \citenamefont {Kanatzidis},\ and\ \citenamefont
		{He}}]{5GeTe}%
	\BibitemOpen
	\bibfield  {author} {\bibinfo {author} {\bibfnamefont {D.}~\bibnamefont
			{Wu}}, \bibinfo {author} {\bibfnamefont {L.-D.}\ \bibnamefont {Zhao}},
		\bibinfo {author} {\bibfnamefont {S.}~\bibnamefont {Hao}}, \bibinfo {author}
		{\bibfnamefont {Q.}~\bibnamefont {Jiang}}, \bibinfo {author} {\bibfnamefont
			{F.}~\bibnamefont {Zheng}}, \bibinfo {author} {\bibfnamefont {J.~W.}\
			\bibnamefont {Doak}}, \bibinfo {author} {\bibfnamefont {H.}~\bibnamefont
			{Wu}}, \bibinfo {author} {\bibfnamefont {H.}~\bibnamefont {Chi}}, \bibinfo
		{author} {\bibfnamefont {Y.}~\bibnamefont {Gelbstein}}, \bibinfo {author}
		{\bibfnamefont {C.}~\bibnamefont {Uher}}, \bibinfo {author} {\bibfnamefont
			{C.}~\bibnamefont {Wolverton}}, \bibinfo {author} {\bibfnamefont
			{M.}~\bibnamefont {Kanatzidis}},\ and\ \bibinfo {author} {\bibfnamefont
			{J.}~\bibnamefont {He}},\ }\bibfield  {title} {\bibinfo {title} {Origin of
			the high performance in {GeTe}-based thermoelectric materials upon
			{Bi$_2$Te$_3$} doping},\ }\href {https://doi.org/10.1021/ja504896a}
	{\bibfield  {journal} {\bibinfo  {journal} {J. Am. Chem. Soc.}\ }\textbf
		{\bibinfo {volume} {136}},\ \bibinfo {pages} {11412} (\bibinfo {year}
		{2014})}\BibitemShut {NoStop}%
	\bibitem [{\citenamefont {Yu}\ \emph {et~al.}(2012)\citenamefont {Yu},
		\citenamefont {Zebarjadi}, \citenamefont {Wang}, \citenamefont {Lukas},
		\citenamefont {Wang}, \citenamefont {Wang}, \citenamefont {Opeil},
		\citenamefont {Dresselhaus}, \citenamefont {Chen},\ and\ \citenamefont
		{Ren}}]{6SiGe}%
	\BibitemOpen
	\bibfield  {author} {\bibinfo {author} {\bibfnamefont {B.}~\bibnamefont
			{Yu}}, \bibinfo {author} {\bibfnamefont {M.}~\bibnamefont {Zebarjadi}},
		\bibinfo {author} {\bibfnamefont {H.}~\bibnamefont {Wang}}, \bibinfo {author}
		{\bibfnamefont {K.}~\bibnamefont {Lukas}}, \bibinfo {author} {\bibfnamefont
			{H.}~\bibnamefont {Wang}}, \bibinfo {author} {\bibfnamefont {D.}~\bibnamefont
			{Wang}}, \bibinfo {author} {\bibfnamefont {C.}~\bibnamefont {Opeil}},
		\bibinfo {author} {\bibfnamefont {M.}~\bibnamefont {Dresselhaus}}, \bibinfo
		{author} {\bibfnamefont {G.}~\bibnamefont {Chen}},\ and\ \bibinfo {author}
		{\bibfnamefont {Z.}~\bibnamefont {Ren}},\ }\bibfield  {title} {\bibinfo
		{title} {Enhancement of thermoelectric properties by modulation-doping in
			silicon germanium alloy nanocomposites},\ }\href
	{https://doi.org/10.1021/nl3003045} {\bibfield  {journal} {\bibinfo
			{journal} {Nano Lett.}\ }\textbf {\bibinfo {volume} {12}},\ \bibinfo {pages}
		{2077} (\bibinfo {year} {2012})}\BibitemShut {NoStop}%
	\bibitem [{\citenamefont {Xin}\ \emph {et~al.}(2018)\citenamefont {Xin},
		\citenamefont {Tang}, \citenamefont {Liu}, \citenamefont {Zhao},
		\citenamefont {Pan},\ and\ \citenamefont {Zhu}}]{7HH1}%
	\BibitemOpen
	\bibfield  {author} {\bibinfo {author} {\bibfnamefont {J.}~\bibnamefont
			{Xin}}, \bibinfo {author} {\bibfnamefont {Y.}~\bibnamefont {Tang}}, \bibinfo
		{author} {\bibfnamefont {Y.}~\bibnamefont {Liu}}, \bibinfo {author}
		{\bibfnamefont {X.}~\bibnamefont {Zhao}}, \bibinfo {author} {\bibfnamefont
			{H.}~\bibnamefont {Pan}},\ and\ \bibinfo {author} {\bibfnamefont
			{T.}~\bibnamefont {Zhu}},\ }\bibfield  {title} {\bibinfo {title}
		{Valleytronics in thermoelectric materials},\ }\href
	{https://doi.org/10.1038/s41535-018-0083-6} {\bibfield  {journal} {\bibinfo
			{journal} {npj Quantum Mater.}\ }\textbf {\bibinfo {volume} {3}},\ \bibinfo
		{eid} {9} (\bibinfo {year} {2018})}\BibitemShut {NoStop}%
	\bibitem [{\citenamefont {Fu}\ \emph {et~al.}(2015)\citenamefont {Fu},
		\citenamefont {Bai}, \citenamefont {Liu}, \citenamefont {Tang}, \citenamefont
		{Chen}, \citenamefont {Zhao},\ and\ \citenamefont {Zhu}}]{8HH2}%
	\BibitemOpen
	\bibfield  {author} {\bibinfo {author} {\bibfnamefont {C.}~\bibnamefont
			{Fu}}, \bibinfo {author} {\bibfnamefont {S.}~\bibnamefont {Bai}}, \bibinfo
		{author} {\bibfnamefont {Y.}~\bibnamefont {Liu}}, \bibinfo {author}
		{\bibfnamefont {Y.}~\bibnamefont {Tang}}, \bibinfo {author} {\bibfnamefont
			{L.}~\bibnamefont {Chen}}, \bibinfo {author} {\bibfnamefont {X.}~\bibnamefont
			{Zhao}},\ and\ \bibinfo {author} {\bibfnamefont {T.}~\bibnamefont {Zhu}},\
	}\bibfield  {title} {\bibinfo {title} {Realizing high figure of merit in
			heavy-band p-type half-heusler thermoelectric materials},\ }\href@noop {}
	{\bibfield  {journal} {\bibinfo  {journal} {Nat. Commun.}\ }\textbf {\bibinfo
			{volume} {6}},\ \bibinfo {pages} {8144} (\bibinfo {year} {2015})}\BibitemShut
	{NoStop}%
	\bibitem [{\citenamefont {Yan}\ \emph {et~al.}()\citenamefont {Yan},
		\citenamefont {Liu}, \citenamefont {Chen}, \citenamefont {Wang},
		\citenamefont {Zhang}, \citenamefont {Chen},\ and\ \citenamefont
		{Ren}}]{9HH3}%
	\BibitemOpen
	\bibfield  {author} {\bibinfo {author} {\bibfnamefont {X.}~\bibnamefont
			{Yan}}, \bibinfo {author} {\bibfnamefont {W.}~\bibnamefont {Liu}}, \bibinfo
		{author} {\bibfnamefont {S.}~\bibnamefont {Chen}}, \bibinfo {author}
		{\bibfnamefont {H.}~\bibnamefont {Wang}}, \bibinfo {author} {\bibfnamefont
			{Q.}~\bibnamefont {Zhang}}, \bibinfo {author} {\bibfnamefont
			{G.}~\bibnamefont {Chen}},\ and\ \bibinfo {author} {\bibfnamefont
			{Z.}~\bibnamefont {Ren}},\ }\bibfield  {title} {\bibinfo {title}
		{Thermoelectric property study of nanostructured p-type half-heuslers {(Hf,
				Zr, Ti)CoSb$_{0.8}$Sn$_{0.2}$}},\ }\href
	{https://doi.org/https://doi.org/10.1002/aenm.201200973} {\bibfield
		{journal} {\bibinfo  {journal} {Adv. Energy Mater.}\ }\textbf {\bibinfo
			{volume} {3}},\ \bibinfo {pages} {1195}}\BibitemShut {NoStop}%
	\bibitem [{\citenamefont {Brod}\ \emph {et~al.}(2022)\citenamefont {Brod},
		\citenamefont {Guo}, \citenamefont {Zhang},\ and\ \citenamefont
		{Snyder}}]{10HH4}%
	\BibitemOpen
	\bibfield  {author} {\bibinfo {author} {\bibfnamefont {M.~K.}\ \bibnamefont
			{Brod}}, \bibinfo {author} {\bibfnamefont {S.}~\bibnamefont {Guo}}, \bibinfo
		{author} {\bibfnamefont {Y.}~\bibnamefont {Zhang}},\ and\ \bibinfo {author}
		{\bibfnamefont {G.~J.}\ \bibnamefont {Snyder}},\ }\bibfield  {title}
	{\bibinfo {title} {Explaining the electronic band structure of half-heusler
			thermoelectric semiconductors for engineering high valley degeneracy},\
	}\href {https://doi.org/10.1557/s43577-022-00360-z} {\bibfield  {journal}
		{\bibinfo  {journal} {MRS Bull.}\ }\textbf {\bibinfo {volume} {47}},\
		\bibinfo {pages} {573–583} (\bibinfo {year} {2022})}\BibitemShut {NoStop}%
	\bibitem [{\citenamefont {Chandra}\ \emph {et~al.}(2022)\citenamefont
		{Chandra}, \citenamefont {Bhat}, \citenamefont {Dutta}, \citenamefont
		{Bhardwaj}, \citenamefont {Datta},\ and\ \citenamefont {Biswas}}]{2022_SnSe}%
	\BibitemOpen
	\bibfield  {author} {\bibinfo {author} {\bibfnamefont {S.}~\bibnamefont
			{Chandra}}, \bibinfo {author} {\bibfnamefont {U.}~\bibnamefont {Bhat}},
		\bibinfo {author} {\bibfnamefont {P.}~\bibnamefont {Dutta}}, \bibinfo
		{author} {\bibfnamefont {A.}~\bibnamefont {Bhardwaj}}, \bibinfo {author}
		{\bibfnamefont {R.}~\bibnamefont {Datta}},\ and\ \bibinfo {author}
		{\bibfnamefont {K.}~\bibnamefont {Biswas}},\ }\bibfield  {title} {\bibinfo
		{title} {Modular nanostructures facilitate low thermal conductivity and
			ultra-high thermoelectric performance in n-type {SnSe}},\ }\href
	{https://doi.org/10.1002/adma.202203725} {\bibfield  {journal} {\bibinfo
			{journal} {Adv. Mater.}\ }\textbf {\bibinfo {volume} {34}},\ \bibinfo {pages}
		{2203725} (\bibinfo {year} {2022})}\BibitemShut {NoStop}%
	\bibitem [{\citenamefont {Vogel}\ and\ \citenamefont
		{Schuster}(1979)}]{Sc1979}%
	\BibitemOpen
	\bibfield  {author} {\bibinfo {author} {\bibfnamefont {R.}~\bibnamefont
			{Vogel}}\ and\ \bibinfo {author} {\bibfnamefont {H.-U.}\ \bibnamefont
			{Schuster}},\ }\bibfield  {title} {\bibinfo {title} {Neue elektrovalente
			ternäre verbindungen des kaliums mit magnesium und elementen der 5.
			hauptgruppe / new ternary compounds of potassium with magnesium and elements
			of the 5. main group},\ }\href {https://doi.org/10.1515/znb-1979-1219}
	{\bibfield  {journal} {\bibinfo  {journal} {Zeitschrift für Naturforschung
				B}\ }\textbf {\bibinfo {volume} {34}},\ \bibinfo {pages} {1719} (\bibinfo
		{year} {1979})}\BibitemShut {NoStop}%
	\bibitem [{\citenamefont {Le}\ \emph {et~al.}(2017)\citenamefont {Le},
		\citenamefont {Qin}, \citenamefont {Wu}, \citenamefont {Dai}, \citenamefont
		{Fu}, \citenamefont {Fang},\ and\ \citenamefont {Hu}}]{KMgBi3}%
	\BibitemOpen
	\bibfield  {author} {\bibinfo {author} {\bibfnamefont {C.}~\bibnamefont
			{Le}}, \bibinfo {author} {\bibfnamefont {S.}~\bibnamefont {Qin}}, \bibinfo
		{author} {\bibfnamefont {X.}~\bibnamefont {Wu}}, \bibinfo {author}
		{\bibfnamefont {X.}~\bibnamefont {Dai}}, \bibinfo {author} {\bibfnamefont
			{P.}~\bibnamefont {Fu}}, \bibinfo {author} {\bibfnamefont {C.}~\bibnamefont
			{Fang}},\ and\ \bibinfo {author} {\bibfnamefont {J.}~\bibnamefont {Hu}},\
	}\bibfield  {title} {\bibinfo {title} {Three-dimensional topological critical
			dirac semimetal in {AMgBi (A= K, Rb, Cs)}},\ }\href
	{https://doi.org/10.1103/PHYSREVB.96.115121/FIGURES/10/MEDIUM} {\bibfield
		{journal} {\bibinfo  {journal} {Phys. Rev. B}\ }\textbf {\bibinfo {volume}
			{96}},\ \bibinfo {pages} {115121} (\bibinfo {year} {2017})}\BibitemShut
	{NoStop}%
	\bibitem [{\citenamefont {Vikram}\ \emph {et~al.}(2022)\citenamefont {Vikram},
		\citenamefont {Sahni}, \citenamefont {Jain},\ and\ \citenamefont
		{Alam}}]{KMgBi2}%
	\BibitemOpen
	\bibfield  {author} {\bibinfo {author} {\bibnamefont {Vikram}}, \bibinfo
		{author} {\bibfnamefont {B.}~\bibnamefont {Sahni}}, \bibinfo {author}
		{\bibfnamefont {A.}~\bibnamefont {Jain}},\ and\ \bibinfo {author}
		{\bibfnamefont {A.}~\bibnamefont {Alam}},\ }\bibfield  {title} {\bibinfo
		{title} {Quasi-2d carrier transport in {KMgBi} for promising thermoelectric
			performance},\ }\href {https://doi.org/10.1021/acsaem.2c01685} {\bibfield
		{journal} {\bibinfo  {journal} {ACS Appl. Energy Mater.}\ }\textbf {\bibinfo
			{volume} {5}},\ \bibinfo {pages} {9141} (\bibinfo {year} {2022})}\BibitemShut
	{NoStop}%
	\bibitem [{\citenamefont {Zhang}\ \emph {et~al.}(2017)\citenamefont {Zhang},
		\citenamefont {Sun},\ and\ \citenamefont {Lei}}]{KMgBi}%
	\BibitemOpen
	\bibfield  {author} {\bibinfo {author} {\bibfnamefont {X.}~\bibnamefont
			{Zhang}}, \bibinfo {author} {\bibfnamefont {S.}~\bibnamefont {Sun}},\ and\
		\bibinfo {author} {\bibfnamefont {H.}~\bibnamefont {Lei}},\ }\bibfield
	{title} {\bibinfo {title} {Narrow-gap semiconducting properties of {KMgBi}
			with multiband feature},\ }\href
	{https://doi.org/10.1103/PHYSREVB.95.035209/FIGURES/5/MEDIUM} {\bibfield
		{journal} {\bibinfo  {journal} {Phys. Rev. B}\ }\textbf {\bibinfo {volume}
			{95}},\ \bibinfo {pages} {035209} (\bibinfo {year} {2017})}\BibitemShut
	{NoStop}%
	\bibitem [{\citenamefont {Bordbar}\ \emph {et~al.}(2022)\citenamefont
		{Bordbar}, \citenamefont {Nedaee-Shakarab},\ and\ \citenamefont
		{Khouzani}}]{KMgP_HH}%
	\BibitemOpen
	\bibfield  {author} {\bibinfo {author} {\bibfnamefont {P.}~\bibnamefont
			{Bordbar}}, \bibinfo {author} {\bibfnamefont {B.}~\bibnamefont
			{Nedaee-Shakarab}},\ and\ \bibinfo {author} {\bibfnamefont {S.~M.}\
			\bibnamefont {Khouzani}},\ }\bibfield  {title} {\bibinfo {title}
		{Thermodynamic phase diagram stability, electronic and thermoelectric
			properties of the half-heusler kmgp {[}111] films},\ }\href
	{https://doi.org/10.1007/s12648-020-01964-4} {\bibfield  {journal} {\bibinfo
			{journal} {Indian J. Phys.}\ }\textbf {\bibinfo {volume} {96}},\ \bibinfo
		{pages} {103} (\bibinfo {year} {2022})}\BibitemShut {NoStop}%
	\bibitem [{\citenamefont {Liu}\ \emph {et~al.}(2021)\citenamefont {Liu},
		\citenamefont {Liu}, \citenamefont {Xing}, \citenamefont {Jiang},\ and\
		\citenamefont {Zhao}}]{KMgP_MoS2_2D}%
	\BibitemOpen
	\bibfield  {author} {\bibinfo {author} {\bibfnamefont {Q.}~\bibnamefont
			{Liu}}, \bibinfo {author} {\bibfnamefont {Y.}~\bibnamefont {Liu}}, \bibinfo
		{author} {\bibfnamefont {J.}~\bibnamefont {Xing}}, \bibinfo {author}
		{\bibfnamefont {X.}~\bibnamefont {Jiang}},\ and\ \bibinfo {author}
		{\bibfnamefont {J.}~\bibnamefont {Zhao}},\ }\bibfield  {title} {\bibinfo
		{title} {A valence balancing rule for the design of bimetallic phosphides
			targeting high thermoelectric performance},\ }\href
	{https://doi.org/10.1039/d1cp02923a} {\bibfield  {journal} {\bibinfo
			{journal} {Phys. Chem. Chem. Phys.}\ }\textbf {\bibinfo {volume} {23}},\
		\bibinfo {pages} {18916} (\bibinfo {year} {2021})}\BibitemShut {NoStop}%
	\bibitem [{\citenamefont {Bennett}\ \emph {et~al.}(2012)\citenamefont
		{Bennett}, \citenamefont {Garrity}, \citenamefont {Rabe},\ and\ \citenamefont
		{Vanderbilt}}]{KMgSb_Hexa}%
	\BibitemOpen
	\bibfield  {author} {\bibinfo {author} {\bibfnamefont {J.~W.}\ \bibnamefont
			{Bennett}}, \bibinfo {author} {\bibfnamefont {K.~F.}\ \bibnamefont
			{Garrity}}, \bibinfo {author} {\bibfnamefont {K.~M.}\ \bibnamefont {Rabe}},\
		and\ \bibinfo {author} {\bibfnamefont {D.}~\bibnamefont {Vanderbilt}},\
	}\bibfield  {title} {\bibinfo {title} {Hexagonal {ABC} semiconductors as
			ferroelectrics},\ }\href {https://doi.org/10.1103/PhysRevLett.109.167602}
	{\bibfield  {journal} {\bibinfo  {journal} {Phys. Rev. Lett.}\ }\textbf
		{\bibinfo {volume} {109}},\ \bibinfo {pages} {167602} (\bibinfo {year}
		{2012})}\BibitemShut {NoStop}%
	\bibitem [{\citenamefont {Bennett}\ \emph {et~al.}(2013)\citenamefont
		{Bennett}, \citenamefont {Garrity}, \citenamefont {Rabe},\ and\ \citenamefont
		{Vanderbilt}}]{KMgSb_Otho}%
	\BibitemOpen
	\bibfield  {author} {\bibinfo {author} {\bibfnamefont {J.~W.}\ \bibnamefont
			{Bennett}}, \bibinfo {author} {\bibfnamefont {K.~F.}\ \bibnamefont
			{Garrity}}, \bibinfo {author} {\bibfnamefont {K.~M.}\ \bibnamefont {Rabe}},\
		and\ \bibinfo {author} {\bibfnamefont {D.}~\bibnamefont {Vanderbilt}},\
	}\bibfield  {title} {\bibinfo {title} {Orthorhombic {ABC} semiconductors as
			antiferroelectrics},\ }\href {https://doi.org/10.1103/PhysRevLett.110.017603}
	{\bibfield  {journal} {\bibinfo  {journal} {Phys. Rev. Lett.}\ }\textbf
		{\bibinfo {volume} {110}},\ \bibinfo {pages} {017603} (\bibinfo {year}
		{2013})}\BibitemShut {NoStop}%
	\bibitem [{\citenamefont {Arif}\ \emph {et~al.}(2016)\citenamefont {Arif},
		\citenamefont {Murtaza}, \citenamefont {Ali}, \citenamefont {Khenata},
		\citenamefont {Takagiwa}, \citenamefont {Muzammil},\ and\ \citenamefont
		{Omran}}]{KMgX_HH}%
	\BibitemOpen
	\bibfield  {author} {\bibinfo {author} {\bibfnamefont {M.}~\bibnamefont
			{Arif}}, \bibinfo {author} {\bibfnamefont {G.}~\bibnamefont {Murtaza}},
		\bibinfo {author} {\bibfnamefont {R.}~\bibnamefont {Ali}}, \bibinfo {author}
		{\bibfnamefont {R.}~\bibnamefont {Khenata}}, \bibinfo {author} {\bibfnamefont
			{Y.}~\bibnamefont {Takagiwa}}, \bibinfo {author} {\bibfnamefont
			{M.}~\bibnamefont {Muzammil}},\ and\ \bibinfo {author} {\bibfnamefont
			{S.~B.}\ \bibnamefont {Omran}},\ }\bibfield  {title} {\bibinfo {title}
		{Elastic and electro-optical properties of {XYZ (X = Li, Na and K; Y = Mg; Z
				= N, P, As, Sb and Bi)} compounds},\ }\href
	{https://doi.org/10.1007/s12648-015-0791-8} {\bibfield  {journal} {\bibinfo
			{journal} {Indian J. Phys.}\ }\textbf {\bibinfo {volume} {90}},\ \bibinfo
		{pages} {639} (\bibinfo {year} {2016})}\BibitemShut {NoStop}%
	\bibitem [{\citenamefont {Xiao}\ and\ \citenamefont {Zhao}(2018)}]{PbTe}%
	\BibitemOpen
	\bibfield  {author} {\bibinfo {author} {\bibfnamefont {Y.}~\bibnamefont
			{Xiao}}\ and\ \bibinfo {author} {\bibfnamefont {L.~D.}\ \bibnamefont
			{Zhao}},\ }\bibfield  {title} {\bibinfo {title} {Charge and phonon transport
			in pbte-based thermoelectric materials},\ }\href
	{https://doi.org/10.1038/s41535-018-0127-y} {\bibfield  {journal} {\bibinfo
			{journal} {npj Quantum Mater.}\ }\textbf {\bibinfo {volume} {3}},\ \bibinfo
		{pages} {1} (\bibinfo {year} {2018})}\BibitemShut {NoStop}%
	\bibitem [{\citenamefont {Fu}\ \emph {et~al.}(2022)\citenamefont {Fu},
		\citenamefont {Zhang}, \citenamefont {Hu}, \citenamefont {Jiang},
		\citenamefont {Huang}, \citenamefont {Ai}, \citenamefont {Wan}, \citenamefont
		{Reith}, \citenamefont {Wang}, \citenamefont {Nielsch},\ and\ \citenamefont
		{Jiang}}]{2022_Mg3Bi2}%
	\BibitemOpen
	\bibfield  {author} {\bibinfo {author} {\bibfnamefont {Y.}~\bibnamefont
			{Fu}}, \bibinfo {author} {\bibfnamefont {Q.}~\bibnamefont {Zhang}}, \bibinfo
		{author} {\bibfnamefont {Z.}~\bibnamefont {Hu}}, \bibinfo {author}
		{\bibfnamefont {M.}~\bibnamefont {Jiang}}, \bibinfo {author} {\bibfnamefont
			{A.}~\bibnamefont {Huang}}, \bibinfo {author} {\bibfnamefont
			{X.}~\bibnamefont {Ai}}, \bibinfo {author} {\bibfnamefont {S.}~\bibnamefont
			{Wan}}, \bibinfo {author} {\bibfnamefont {H.}~\bibnamefont {Reith}}, \bibinfo
		{author} {\bibfnamefont {L.}~\bibnamefont {Wang}}, \bibinfo {author}
		{\bibfnamefont {K.}~\bibnamefont {Nielsch}},\ and\ \bibinfo {author}
		{\bibfnamefont {W.}~\bibnamefont {Jiang}},\ }\bibfield  {title} {\bibinfo
		{title} {{Mg$_3$(Bi{,}Sb)$_2$}-based thermoelectric modules for efficient and
			reliable waste-heat utilization up to {750 K}},\ }\href
	{https://doi.org/10.1039/D2EE01038K} {\bibfield  {journal} {\bibinfo
			{journal} {Energy Environ. Sci.}\ }\textbf {\bibinfo {volume} {15}},\
		\bibinfo {pages} {3265} (\bibinfo {year} {2022})}\BibitemShut {NoStop}%
	\bibitem [{\citenamefont {Kresse}\ and\ \citenamefont {Hafner}(1993)}]{1VASP}%
	\BibitemOpen
	\bibfield  {author} {\bibinfo {author} {\bibfnamefont {G.}~\bibnamefont
			{Kresse}}\ and\ \bibinfo {author} {\bibfnamefont {J.}~\bibnamefont
			{Hafner}},\ }\bibfield  {title} {\bibinfo {title} {Ab initio molecular
			dynamics for liquid metals},\ }\href
	{https://doi.org/10.1103/PhysRevB.47.558} {\bibfield  {journal} {\bibinfo
			{journal} {Phys. Rev. B}\ }\textbf {\bibinfo {volume} {47}},\ \bibinfo
		{pages} {558} (\bibinfo {year} {1993})}\BibitemShut {NoStop}%
	\bibitem [{\citenamefont {Kresse}\ and\ \citenamefont {Hafner}(1994)}]{2VASP}%
	\BibitemOpen
	\bibfield  {author} {\bibinfo {author} {\bibfnamefont {G.}~\bibnamefont
			{Kresse}}\ and\ \bibinfo {author} {\bibfnamefont {J.}~\bibnamefont
			{Hafner}},\ }\bibfield  {title} {\bibinfo {title} {Ab initio
			molecular-dynamics simulation of the liquid-metal--amorphous-semiconductor
			transition in germanium},\ }\href {https://doi.org/10.1103/PhysRevB.49.14251}
	{\bibfield  {journal} {\bibinfo  {journal} {Phys. Rev. B}\ }\textbf {\bibinfo
			{volume} {49}},\ \bibinfo {pages} {14251} (\bibinfo {year}
		{1994})}\BibitemShut {NoStop}%
	\bibitem [{\citenamefont {Kressse}\ and\ \citenamefont
		{Furthm{\"u}ller}(1996)}]{3VASP}%
	\BibitemOpen
	\bibfield  {author} {\bibinfo {author} {\bibfnamefont {G.}~\bibnamefont
			{Kressse}}\ and\ \bibinfo {author} {\bibfnamefont {J.}~\bibnamefont
			{Furthm{\"u}ller}},\ }\bibfield  {title} {\bibinfo {title} {Efficiency of
			ab-initio total energy calculations for metals and semiconductors using a
			plane-wave basis set},\ }\href {https://doi.org/10.1016/0927-0256(96)00008-0}
	{\bibfield  {journal} {\bibinfo  {journal} {Comput. Mater. Sci.}\ }\textbf
		{\bibinfo {volume} {6}},\ \bibinfo {pages} {15} (\bibinfo {year}
		{1996})}\BibitemShut {NoStop}%
	\bibitem [{\citenamefont {Kresse}\ and\ \citenamefont
		{Furthm{\"u}ller}(1996)}]{4VASP}%
	\BibitemOpen
	\bibfield  {author} {\bibinfo {author} {\bibfnamefont {G.}~\bibnamefont
			{Kresse}}\ and\ \bibinfo {author} {\bibfnamefont {J.}~\bibnamefont
			{Furthm{\"u}ller}},\ }\bibfield  {title} {\bibinfo {title} {Efficient
			iterative schemes for ab initio total-energy calculations using a plane-wave
			basis set},\ }\href {https://doi.org/10.1103/PhysRevB.54.11169} {\bibfield
		{journal} {\bibinfo  {journal} {Phys. Rev. B}\ }\textbf {\bibinfo {volume}
			{54}},\ \bibinfo {pages} {11169} (\bibinfo {year} {1996})}\BibitemShut
	{NoStop}%
	\bibitem [{\citenamefont {Kresse}\ and\ \citenamefont {Joubert}(1999)}]{5VASP}%
	\BibitemOpen
	\bibfield  {author} {\bibinfo {author} {\bibfnamefont {G.}~\bibnamefont
			{Kresse}}\ and\ \bibinfo {author} {\bibfnamefont {D.}~\bibnamefont
			{Joubert}},\ }\bibfield  {title} {\bibinfo {title} {From ultrasoft
			pseudopotentials to the projector augmented-wave method},\ }\href
	{https://doi.org/10.1103/PhysRevB.59.1758} {\bibfield  {journal} {\bibinfo
			{journal} {Phys. Rev. B}\ }\textbf {\bibinfo {volume} {59}},\ \bibinfo
		{pages} {1758} (\bibinfo {year} {1999})}\BibitemShut {NoStop}%
	\bibitem [{\citenamefont {Bl\"ochl}(1994)}]{paw1}%
	\BibitemOpen
	\bibfield  {author} {\bibinfo {author} {\bibfnamefont {P.~E.}\ \bibnamefont
			{Bl\"ochl}},\ }\bibfield  {title} {\bibinfo {title} {Projector augmented-wave
			method},\ }\href {https://doi.org/10.1103/PhysRevB.50.17953} {\bibfield
		{journal} {\bibinfo  {journal} {Phys. Rev. B}\ }\textbf {\bibinfo {volume}
			{50}},\ \bibinfo {pages} {17953} (\bibinfo {year} {1994})}\BibitemShut
	{NoStop}%
	\bibitem [{\citenamefont {Perdew}\ \emph {et~al.}(1996)\citenamefont {Perdew},
		\citenamefont {Burke},\ and\ \citenamefont {Ernzerhof}}]{GGA}%
	\BibitemOpen
	\bibfield  {author} {\bibinfo {author} {\bibfnamefont {J.~P.}\ \bibnamefont
			{Perdew}}, \bibinfo {author} {\bibfnamefont {K.}~\bibnamefont {Burke}},\ and\
		\bibinfo {author} {\bibfnamefont {M.}~\bibnamefont {Ernzerhof}},\ }\bibfield
	{title} {\bibinfo {title} {Generalized gradient approximation made simple},\
	}\href {https://doi.org/10.1103/PhysRevLett.77.3865} {\bibfield  {journal}
		{\bibinfo  {journal} {Phys. Rev. Lett.}\ }\textbf {\bibinfo {volume} {77}},\
		\bibinfo {pages} {3865} (\bibinfo {year} {1996})}\BibitemShut {NoStop}%
	\bibitem [{\citenamefont {Perdew}\ \emph {et~al.}(2008)\citenamefont {Perdew},
		\citenamefont {Ruzsinszky}, \citenamefont {Csonka}, \citenamefont {Vydrov},
		\citenamefont {Scuseria}, \citenamefont {Constantin}, \citenamefont {Zhou},\
		and\ \citenamefont {Burke}}]{PBEsol}%
	\BibitemOpen
	\bibfield  {author} {\bibinfo {author} {\bibfnamefont {J.~P.}\ \bibnamefont
			{Perdew}}, \bibinfo {author} {\bibfnamefont {A.}~\bibnamefont {Ruzsinszky}},
		\bibinfo {author} {\bibfnamefont {G.~I.}\ \bibnamefont {Csonka}}, \bibinfo
		{author} {\bibfnamefont {O.~A.}\ \bibnamefont {Vydrov}}, \bibinfo {author}
		{\bibfnamefont {G.~E.}\ \bibnamefont {Scuseria}}, \bibinfo {author}
		{\bibfnamefont {L.~A.}\ \bibnamefont {Constantin}}, \bibinfo {author}
		{\bibfnamefont {X.}~\bibnamefont {Zhou}},\ and\ \bibinfo {author}
		{\bibfnamefont {K.}~\bibnamefont {Burke}},\ }\bibfield  {title} {\bibinfo
		{title} {Restoring the density-gradient expansion for exchange in solids and
			surfaces},\ }\href {https://doi.org/10.1103/PhysRevLett.100.136406}
	{\bibfield  {journal} {\bibinfo  {journal} {Phys. Rev. Lett.}\ }\textbf
		{\bibinfo {volume} {100}},\ \bibinfo {pages} {136406} (\bibinfo {year}
		{2008})}\BibitemShut {NoStop}%
	\bibitem [{\citenamefont {Krukau}\ \emph {et~al.}(2006)\citenamefont {Krukau},
		\citenamefont {Vydrov}, \citenamefont {Izmaylov},\ and\ \citenamefont
		{Scuseria}}]{HSE06}%
	\BibitemOpen
	\bibfield  {author} {\bibinfo {author} {\bibfnamefont {A.~V.}\ \bibnamefont
			{Krukau}}, \bibinfo {author} {\bibfnamefont {O.~A.}\ \bibnamefont {Vydrov}},
		\bibinfo {author} {\bibfnamefont {A.~F.}\ \bibnamefont {Izmaylov}},\ and\
		\bibinfo {author} {\bibfnamefont {G.~E.}\ \bibnamefont {Scuseria}},\
	}\bibfield  {title} {\bibinfo {title} {Influence of the exchange screening
			parameter on the performance of screened hybrid functionals},\ }\href
	{https://doi.org/10.1063/1.2404663} {\bibfield  {journal} {\bibinfo
			{journal} {J. Chem. Phys.}\ }\textbf {\bibinfo {volume} {125}},\ \bibinfo
		{pages} {224106} (\bibinfo {year} {2006})}\BibitemShut {NoStop}%
	\bibitem [{\citenamefont {Klime\ifmmode~\check{s}\else \v{s}\fi{}}\ \emph
		{et~al.}(2011)\citenamefont {Klime\ifmmode~\check{s}\else \v{s}\fi{}},
		\citenamefont {Bowler},\ and\ \citenamefont {Michaelides}}]{OptB86b-vdW}%
	\BibitemOpen
	\bibfield  {author} {\bibinfo {author} {\bibfnamefont {J.~c.~v.}\
			\bibnamefont {Klime\ifmmode~\check{s}\else \v{s}\fi{}}}, \bibinfo {author}
		{\bibfnamefont {D.~R.}\ \bibnamefont {Bowler}},\ and\ \bibinfo {author}
		{\bibfnamefont {A.}~\bibnamefont {Michaelides}},\ }\bibfield  {title}
	{\bibinfo {title} {Van der waals density functionals applied to solids},\
	}\href {https://doi.org/10.1103/PhysRevB.83.195131} {\bibfield  {journal}
		{\bibinfo  {journal} {Phys. Rev. B}\ }\textbf {\bibinfo {volume} {83}},\
		\bibinfo {pages} {195131} (\bibinfo {year} {2011})}\BibitemShut {NoStop}%
	\bibitem [{\citenamefont {Togo}\ and\ \citenamefont {Tanaka}(2015)}]{phonopy}%
	\BibitemOpen
	\bibfield  {author} {\bibinfo {author} {\bibfnamefont {A.}~\bibnamefont
			{Togo}}\ and\ \bibinfo {author} {\bibfnamefont {I.}~\bibnamefont {Tanaka}},\
	}\bibfield  {title} {\bibinfo {title} {First principles phonon calculations
			in materials science},\ }\href
	{https://doi.org/10.1016/j.scriptamat.2015.07.021} {\bibfield  {journal}
		{\bibinfo  {journal} {Scr. Mater.}\ }\textbf {\bibinfo {volume} {108}},\
		\bibinfo {pages} {1} (\bibinfo {year} {2015})}\BibitemShut {NoStop}%
	\bibitem [{\citenamefont {Togo}\ \emph {et~al.}(2015)\citenamefont {Togo},
		\citenamefont {Chaput},\ and\ \citenamefont {Tanaka}}]{phono3py}%
	\BibitemOpen
	\bibfield  {author} {\bibinfo {author} {\bibfnamefont {A.}~\bibnamefont
			{Togo}}, \bibinfo {author} {\bibfnamefont {L.}~\bibnamefont {Chaput}},\ and\
		\bibinfo {author} {\bibfnamefont {I.}~\bibnamefont {Tanaka}},\ }\bibfield
	{title} {\bibinfo {title} {Distributions of phonon lifetimes in brillouin
			zones},\ }\href {https://doi.org/10.1103/PhysRevB.91.094306} {\bibfield
		{journal} {\bibinfo  {journal} {Phys. Rev. B}\ }\textbf {\bibinfo {volume}
			{91}},\ \bibinfo {pages} {094306} (\bibinfo {year} {2015})}\BibitemShut
	{NoStop}%
	\bibitem [{\citenamefont {Li}\ \emph {et~al.}(2014)\citenamefont {Li},
		\citenamefont {Carrete}, \citenamefont {Katcho},\ and\ \citenamefont
		{Mingo}}]{ShengBTE_2014}%
	\BibitemOpen
	\bibfield  {author} {\bibinfo {author} {\bibfnamefont {W.}~\bibnamefont
			{Li}}, \bibinfo {author} {\bibfnamefont {J.}~\bibnamefont {Carrete}},
		\bibinfo {author} {\bibfnamefont {N.~A.}\ \bibnamefont {Katcho}},\ and\
		\bibinfo {author} {\bibfnamefont {N.}~\bibnamefont {Mingo}},\ }\bibfield
	{title} {\bibinfo {title} {{ShengBTE:} a solver of the {B}oltzmann transport
			equation for phonons},\ }\href {https://doi.org/10.1016/j.cpc.2014.02.015}
	{\bibfield  {journal} {\bibinfo  {journal} {Comp. Phys. Commun.}\ }\textbf
		{\bibinfo {volume} {185}},\ \bibinfo {pages} {1747–1758} (\bibinfo {year}
		{2014})}\BibitemShut {NoStop}%
	\bibitem [{\citenamefont {Li}\ \emph {et~al.}(2012)\citenamefont {Li},
		\citenamefont {Lindsay}, \citenamefont {Broido}, \citenamefont {Stewart},\
		and\ \citenamefont {Mingo}}]{thirdorder}%
	\BibitemOpen
	\bibfield  {author} {\bibinfo {author} {\bibfnamefont {W.}~\bibnamefont
			{Li}}, \bibinfo {author} {\bibfnamefont {L.}~\bibnamefont {Lindsay}},
		\bibinfo {author} {\bibfnamefont {D.~A.}\ \bibnamefont {Broido}}, \bibinfo
		{author} {\bibfnamefont {D.~A.}\ \bibnamefont {Stewart}},\ and\ \bibinfo
		{author} {\bibfnamefont {N.}~\bibnamefont {Mingo}},\ }\bibfield  {title}
	{\bibinfo {title} {Thermal conductivity of bulk and nanowire
			{Mg${_2}$Si${_x}$Sn${_{1-x}}$} alloys from first principles},\ }\href@noop {}
	{\bibfield  {journal} {\bibinfo  {journal} {Phys. Rev. B}\ }\textbf {\bibinfo
			{volume} {86}},\ \bibinfo {pages} {174307} (\bibinfo {year}
		{2012})}\BibitemShut {NoStop}%
	\bibitem [{\citenamefont {Feng}\ and\ \citenamefont
		{Ruan}(2016)}]{2016_Four_Phonon}%
	\BibitemOpen
	\bibfield  {author} {\bibinfo {author} {\bibfnamefont {T.}~\bibnamefont
			{Feng}}\ and\ \bibinfo {author} {\bibfnamefont {X.}~\bibnamefont {Ruan}},\
	}\bibfield  {title} {\bibinfo {title} {Quantum mechanical prediction of
			four-phonon scattering rates and reduced thermal conductivity of solids},\
	}\href {https://doi.org/10.1103/PhysRevB.93.045202} {\bibfield  {journal}
		{\bibinfo  {journal} {Phys. Rev. B}\ }\textbf {\bibinfo {volume} {93}},\
		\bibinfo {pages} {045202} (\bibinfo {year} {2016})}\BibitemShut {NoStop}%
	\bibitem [{\citenamefont {Feng}\ \emph {et~al.}(2017)\citenamefont {Feng},
		\citenamefont {Lindsay},\ and\ \citenamefont {Ruan}}]{2017_Four_Phonon}%
	\BibitemOpen
	\bibfield  {author} {\bibinfo {author} {\bibfnamefont {T.}~\bibnamefont
			{Feng}}, \bibinfo {author} {\bibfnamefont {L.}~\bibnamefont {Lindsay}},\ and\
		\bibinfo {author} {\bibfnamefont {X.}~\bibnamefont {Ruan}},\ }\bibfield
	{title} {\bibinfo {title} {Four-phonon scattering significantly reduces
			intrinsic thermal conductivity of solids},\ }\href
	{https://doi.org/10.1103/PhysRevB.96.161201} {\bibfield  {journal} {\bibinfo
			{journal} {Phys. Rev. B}\ }\textbf {\bibinfo {volume} {96}},\ \bibinfo
		{pages} {161201} (\bibinfo {year} {2017})}\BibitemShut {NoStop}%
	\bibitem [{\citenamefont {Han}\ \emph {et~al.}(2022)\citenamefont {Han},
		\citenamefont {Yang}, \citenamefont {Li}, \citenamefont {Feng},\ and\
		\citenamefont {Ruan}}]{2022_Four_Phonon}%
	\BibitemOpen
	\bibfield  {author} {\bibinfo {author} {\bibfnamefont {Z.}~\bibnamefont
			{Han}}, \bibinfo {author} {\bibfnamefont {X.}~\bibnamefont {Yang}}, \bibinfo
		{author} {\bibfnamefont {W.}~\bibnamefont {Li}}, \bibinfo {author}
		{\bibfnamefont {T.}~\bibnamefont {Feng}},\ and\ \bibinfo {author}
		{\bibfnamefont {X.}~\bibnamefont {Ruan}},\ }\bibfield  {title} {\bibinfo
		{title} {Fourphonon: An extension module to shengbte for computing
			four-phonon scattering rates and thermal conductivity},\ }\href
	{https://doi.org/https://doi.org/10.1016/j.cpc.2021.108179} {\bibfield
		{journal} {\bibinfo  {journal} {Comput. Phys. Commun.}\ }\textbf {\bibinfo
			{volume} {270}},\ \bibinfo {pages} {108179} (\bibinfo {year}
		{2022})}\BibitemShut {NoStop}%
	\bibitem [{\citenamefont {Ganose}\ \emph {et~al.}(2021)\citenamefont {Ganose},
		\citenamefont {Park}, \citenamefont {Faghaninia}, \citenamefont
		{Woods-Robinson}, \citenamefont {Persson},\ and\ \citenamefont
		{Jain}}]{AMSET}%
	\BibitemOpen
	\bibfield  {author} {\bibinfo {author} {\bibfnamefont {A.~M.}\ \bibnamefont
			{Ganose}}, \bibinfo {author} {\bibfnamefont {J.}~\bibnamefont {Park}},
		\bibinfo {author} {\bibfnamefont {A.}~\bibnamefont {Faghaninia}}, \bibinfo
		{author} {\bibfnamefont {R.}~\bibnamefont {Woods-Robinson}}, \bibinfo
		{author} {\bibfnamefont {K.~A.}\ \bibnamefont {Persson}},\ and\ \bibinfo
		{author} {\bibfnamefont {A.}~\bibnamefont {Jain}},\ }\bibfield  {title}
	{\bibinfo {title} {Efficient calculation of carrier scattering rates from
			first principles},\ }\href {https://doi.org/10.1038/s41467-021-22440-5}
	{\bibfield  {journal} {\bibinfo  {journal} {Nat. Commun.}\ }\textbf {\bibinfo
			{volume} {12}},\ \bibinfo {pages} {2222} (\bibinfo {year}
		{2021})}\BibitemShut {NoStop}%
	\bibitem [{\citenamefont {Madsen}\ \emph {et~al.}(2018)\citenamefont {Madsen},
		\citenamefont {Carrete},\ and\ \citenamefont {Verstraete}}]{btp2}%
	\BibitemOpen
	\bibfield  {author} {\bibinfo {author} {\bibfnamefont {G.~K.}\ \bibnamefont
			{Madsen}}, \bibinfo {author} {\bibfnamefont {J.}~\bibnamefont {Carrete}},\
		and\ \bibinfo {author} {\bibfnamefont {M.~J.}\ \bibnamefont {Verstraete}},\
	}\bibfield  {title} {\bibinfo {title} {{BoltzTraP2}, a program for
			interpolating band structures and calculating semi-classical transport
			coefficients},\ }\href
	{https://doi.org/https://doi.org/10.1016/j.cpc.2018.05.010} {\bibfield
		{journal} {\bibinfo  {journal} {Comput. Phys. Commun.}\ }\textbf {\bibinfo
			{volume} {231}},\ \bibinfo {pages} {140} (\bibinfo {year}
		{2018})}\BibitemShut {NoStop}%
	\bibitem [{\citenamefont {Onsager}(1931)}]{onsager}%
	\BibitemOpen
	\bibfield  {author} {\bibinfo {author} {\bibfnamefont {L.}~\bibnamefont
			{Onsager}},\ }\bibfield  {title} {\bibinfo {title} {Reciprocal relations in
			irreversible processes. {I.}},\ }\href
	{https://doi.org/10.1103/PhysRev.37.405} {\bibfield  {journal} {\bibinfo
			{journal} {Phys. Rev.}\ }\textbf {\bibinfo {volume} {37}},\ \bibinfo {pages}
		{405} (\bibinfo {year} {1931})}\BibitemShut {NoStop}%
	\bibitem [{\citenamefont {Slack}(1979)}]{slack}%
	\BibitemOpen
	\bibfield  {author} {\bibinfo {author} {\bibfnamefont {G.~A.}\ \bibnamefont
			{Slack}},\ }\href
	{https://doi.org/https://doi.org/10.1016/S0081-1947(08)60359-8} {\emph
		{\bibinfo {title} {The Thermal Conductivity of Nonmetallic Crystals}}},\
	edited by\ \bibinfo {editor} {\bibfnamefont {H.}~\bibnamefont {Ehrenreich}},
	\bibinfo {editor} {\bibfnamefont {F.}~\bibnamefont {Seitz}},\ and\ \bibinfo
	{editor} {\bibfnamefont {D.}~\bibnamefont {Turnbull}},\ \bibinfo {series}
	{Solid State Physics}, Vol.~\bibinfo {volume} {34}\ (\bibinfo  {publisher}
	{Academic Press},\ \bibinfo {year} {1979})\ pp.\ \bibinfo {pages} {1--71},\
	\bibinfo {note} {and references therein}\BibitemShut {NoStop}%
	\bibitem [{\citenamefont {She}\ \emph {et~al.}(2023)\citenamefont {She},
		\citenamefont {Zhao}, \citenamefont {Ni}, \citenamefont {Meng},\ and\
		\citenamefont {Dai}}]{2023_SHE}%
	\BibitemOpen
	\bibfield  {author} {\bibinfo {author} {\bibfnamefont {A.}~\bibnamefont
			{She}}, \bibinfo {author} {\bibfnamefont {Y.}~\bibnamefont {Zhao}}, \bibinfo
		{author} {\bibfnamefont {J.}~\bibnamefont {Ni}}, \bibinfo {author}
		{\bibfnamefont {S.}~\bibnamefont {Meng}},\ and\ \bibinfo {author}
		{\bibfnamefont {Z.}~\bibnamefont {Dai}},\ }\bibfield  {title} {\bibinfo
		{title} {Investigation on transport properties and anomalously heat-carrying
			optical phonons in {KXY (X = Ca, Mg; Y = Sb, Bi)}},\ }\href
	{https://doi.org/10.1016/j.ijheatmasstransfer.2023.124132} {\bibfield
		{journal} {\bibinfo  {journal} {Int. J. Heat Mass Transf.}\ }\textbf
		{\bibinfo {volume} {209}},\ \bibinfo {pages} {124132} (\bibinfo {year}
		{2023})}\BibitemShut {NoStop}%
	\bibitem [{\citenamefont {Tadano}\ \emph {et~al.}(2014)\citenamefont {Tadano},
		\citenamefont {Gohda},\ and\ \citenamefont {Tsuneyuki}}]{Tadano_2014}%
	\BibitemOpen
	\bibfield  {author} {\bibinfo {author} {\bibfnamefont {T.}~\bibnamefont
			{Tadano}}, \bibinfo {author} {\bibfnamefont {Y.}~\bibnamefont {Gohda}},\ and\
		\bibinfo {author} {\bibfnamefont {S.}~\bibnamefont {Tsuneyuki}},\ }\bibfield
	{title} {\bibinfo {title} {Anharmonic force constants extracted from
			first-principles molecular dynamics: applications to heat transfer
			simulations},\ }\href {https://doi.org/10.1088/0953-8984/26/22/225402}
	{\bibfield  {journal} {\bibinfo  {journal} {J. Phys. Condens. Matter}\
		}\textbf {\bibinfo {volume} {26}},\ \bibinfo {pages} {225402} (\bibinfo
		{year} {2014})}\BibitemShut {NoStop}%
	\bibitem [{\citenamefont {Tadano}\ and\ \citenamefont
		{Tsuneyuki}(2015)}]{Tadano_2015}%
	\BibitemOpen
	\bibfield  {author} {\bibinfo {author} {\bibfnamefont {T.}~\bibnamefont
			{Tadano}}\ and\ \bibinfo {author} {\bibfnamefont {S.}~\bibnamefont
			{Tsuneyuki}},\ }\bibfield  {title} {\bibinfo {title} {Self-consistent phonon
			calculations of lattice dynamical properties in cubic {SrTiO}$_{3}$ with
			first-principles anharmonic force constants},\ }\href
	{https://doi.org/10.1103/PhysRevB.92.054301} {\bibfield  {journal} {\bibinfo
			{journal} {Phys. Rev. B}\ }\textbf {\bibinfo {volume} {92}},\ \bibinfo
		{pages} {054301} (\bibinfo {year} {2015})}\BibitemShut {NoStop}%
	\bibitem [{\citenamefont {Singh}\ and\ \citenamefont {Mazin}(1997)}]{CRTA}%
	\BibitemOpen
	\bibfield  {author} {\bibinfo {author} {\bibfnamefont {D.~J.}\ \bibnamefont
			{Singh}}\ and\ \bibinfo {author} {\bibfnamefont {I.~I.}\ \bibnamefont
			{Mazin}},\ }\bibfield  {title} {\bibinfo {title} {Calculated thermoelectric
			properties of {La}-filled skutterudites},\ }\href
	{https://doi.org/10.1103/PhysRevB.56.R1650} {\bibfield  {journal} {\bibinfo
			{journal} {Phys. Rev. B}\ }\textbf {\bibinfo {volume} {56}},\ \bibinfo
		{pages} {R1650} (\bibinfo {year} {1997})}\BibitemShut {NoStop}%
	\bibitem [{sup()}]{supp}%
	\BibitemOpen
	\href@noop {} {}\bibinfo {howpublished}
	{\url{URL_will_be_inserted_by_publisher}}\BibitemShut {NoStop}%
\end{thebibliography}

%

\end{document}